# The Impact of Corona Populism: Empirical Evidence from Austria and Theory

Patrick Mellacher[1]

[1] University of Graz, Graz Schumpeter Centre,

Universitätsstraße 15/FE, A-8010 Graz

patrick.mellacher@uni-graz.at

**Abstract**

I study the co-evolution between public opinion and party policy in situations of crises by investigating a policy U-turn of a major Austrian right-wing party (FPÖ) during the Covid-19 pandemic. My analysis suggests the existence of both i) a "Downsian" effect, which causes voters to adapt their party preferences based on policy congruence and ii) a "party identification" effect, which causes partisans to realign their policy preferences based on "their" party's platform. Specifically, I use individual-level panel data to show that i) "corona skeptical" voters who did not vote for the FPÖ in the pre-Covid-19 elections of 2019 were more likely to vote for the party after it embraced "corona populism", and ii) beliefs of respondents who declared that they voted for the FPÖ in 2019 diverged from the rest of the population in three out of four health-related dimensions only after the turn, causing them to underestimate the threat posed by Covid-19 compared to the rest of the population. Using aggregate-level panel data, I study whether the turn has produced significant behavioral differences which could be observed in terms of reported cases and deaths per capita. Paradoxically, after the turn the FPÖ vote share is significantly positively correlated with deaths per capita, but not with the reported number of infections. I hypothesize that this can be traced back to a self-selection bias in testing, which causes a correlation between the number of "corona skeptics" and the share of unreported cases after the turn. I find empirical support for this hypothesis in individual-level data from a Covid-19 prevalence study that involves information about participants' true vs. reported infection status. I finally study a simple heterogeneous mixing epidemiological model and show that a testing bias can indeed explain the apparent paradox of an increase in deaths without an increase in reported cases. My results can, among others, be used to enrich formal analyses regarding the co-evolution between voter and party behavior.

## 1 Introduction

In situations of severe crises, public opinion and party platforms may shift drastically and frequently. Understanding these opinion dynamics is key to face those crises, as science skepticism may impede evidence-based efforts to tackle them.

A prime example for a rapidly moving crisis has been given by the Covid-19 pandemic, which has compelled governments around the world to enforce strict measures to contain the spread of the virus. After the dust of the first wave of infections has settled, however, skepticism about the necessity of containment measures had been on the rise.

In principle, such skepticism may stem from two sources: compared to the majority of the population, skeptics may either value the benefits of containment measures lower and/or their costs higher. While the benefits of containment measures have mostly been conceived as reducing the number of infections and deaths, the costs seem to be both of a social/psychological (e.g. forgone leisure activities) and an economic nature (supply and/or demand shocks).

Studying trade-offs is a natural habitat for economists, who rushed to study a potential tradeoff between “lives and livelihoods” using coupled economic-epidemiological models, thereby emphasizing the role of economic costs (e.g. Acemoglu et al. 2021; Mendoza et al. 2020; Thunström et al. 2021). It is important to note, however, that not all research done in this tradition agrees upon a clear health-economy trade-off, in particular as a number of authors consider the case of endogenous social distancing, i.e. individuals reacting to the spread of a deadly virus by reducing their social contacts and consumption on their own, which may cause similar or even higher recessions than state-ordered social distancing (see e.g. Aum et al. 2021; Bethune and Korinek 2020; Piguillem and Shi 2022).

Beliefs about a potential trade-off between (economic) costs and (health) benefits may not only be relevant for the public discourse, but also for individual social distancing behavior. This idea was already formalized by Allcott et al. (2020), who consider both differences in *rational perceptions* and *irrational beliefs* between Republicans and Democrats. Further formalizing the idea of behavior under this trade-off, Proaño and Makarewicz (2021) develop a coupled macroeconomic SIR model in which the number of skeptics depends on the current economic and epidemiological situation.

However, skepticism is not confined to the individual level. Once skepticism becomes a political platform, we may call it “corona populism”. Corona populism can be viewed as the political wing of skepticism seeking to i) exploit skepticism, ii) promote it and ii) ultimately implement “skeptical” policies. Economically, this is consistent with Acemoglu et al. (2013), who view populism as a political strategy aimed at exploiting anti-elitist views that receive significant support, but ultimately have adverse effects for the majority of the population. The anti-elitist element inherent to “corona populism”, which opposes the scientific mainstream, is a defining characteristic of populism for other, even more common, definitions of populism (e.g. Hawkins and Rovira Kaltwasser 2018).

One important question regarding the relationship between corona skepticism and corona populism is whether skepticism creates populist politicians and parties, or whether corona populist politics fuel skepticism. While literature in the tradition of Downs (1957) assumes that political programs are set to maximize votes, and are hence determined by voter preferences, other strands in the political science literature (e.g. Campbell et al. 1980), as well as more recent literature drawing on experiments (Bechtel et al. 2015; Druckman et al. 2013; Grewenig et al. 2020) and natural experiments (Slothuus 2010) suggest that voters adapt their policy preferences to match with their party identification.

A more nuanced view on this question would suggest that any important real-world political process exhibits both channels: A “Downsian” effect that drives parties to adopt political platforms that increase their support among the electorate, as well as a “party identification” effect that drives voters who strongly identify with their partisan affiliation to adapt their preferences to match their party’s platform (see fig. 1).

We can view the former effect as a result of a “rational” pursuit of voters to vote for a party that best suits their preferences and a “rational” response from the political parties to this behavior. The latter effect, on the other hand, may be driven, e.g., by a desire to reduce cognitive dissonance, or by viewing the party platform genuinely as an important source for information.

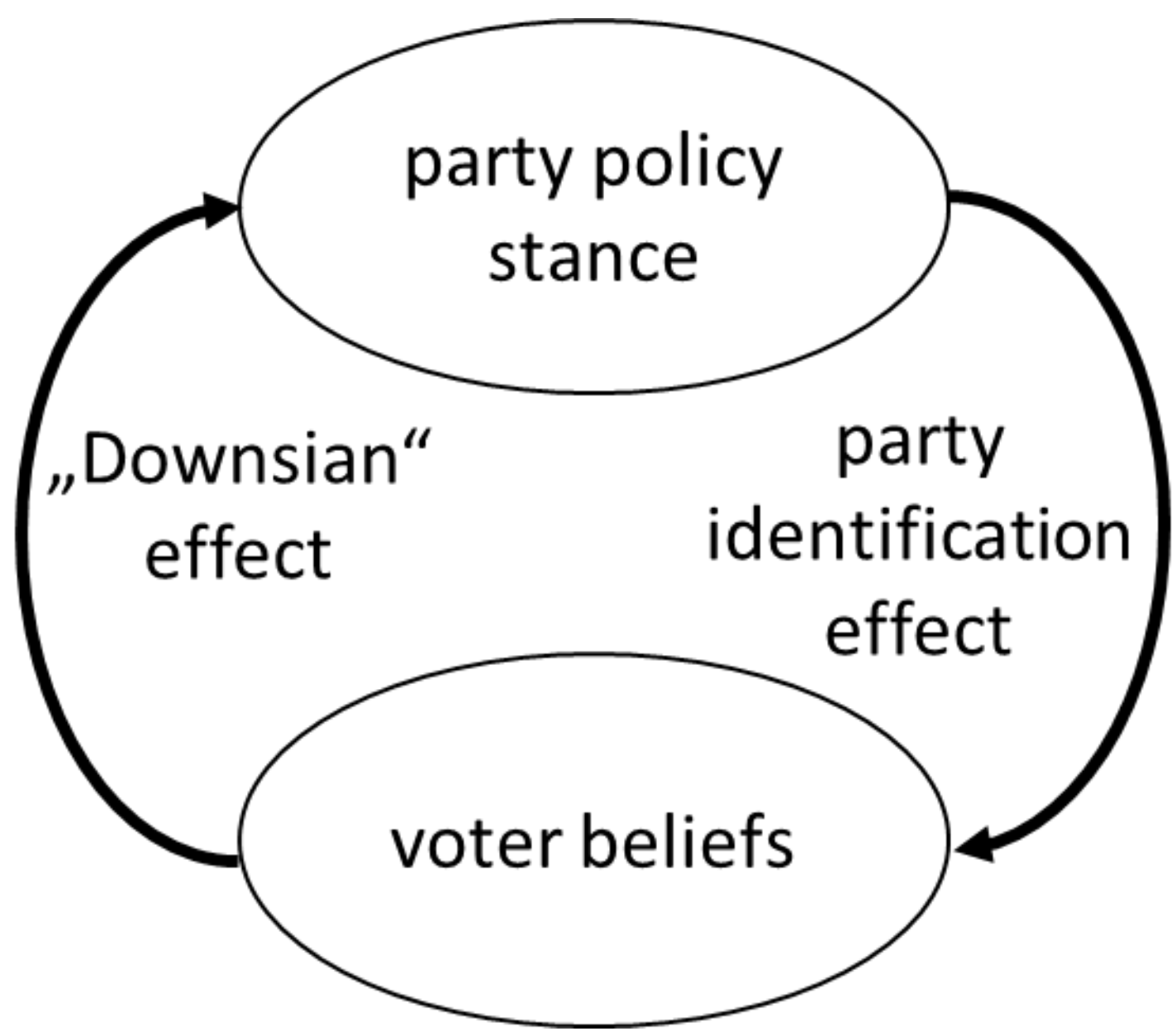

Figure 1: „Downsian" effect vs. "party identification" effect

In this paper, I study a policy U-turn of the Austrian freedom party (FPÖ), a long-established major right-wing populist (e.g. Huber et al. 2021) opposition party, to shed light on the relationship between party corona populism and skepticism specifically and voter preferences and party platform more generally.

The FPÖ was the first political party to demand that the Austrian government should take drastic measures against COVID-19 by demanding a national lockdown on the 13th of March 2020 (APA OTS 2020a). By the end of April of the same year, however, the FPÖ made a U-turn and demanded to "end the Corona madness" (APA OTS 2020b) by which they meant the containment measures taken by the government. In the end of November 2020, one representative of the party even went so far as to advise people not to participate in a mass testing program announced by the Austrian government to be held before Christmas because testing positive would mean that they would have to spend Christmas home alone (APA OTS 2020c). On January 31st 2021, three MPs of the FPÖ participated at a banned demonstration against the lockdown (APA OTS 2021a). Later on, leading FPÖ politicians argued against the introduction of the "green pass" in March 2021 (APA OTS 2021b), as well as against vaccinating teenagers in June 2021 (APA OTS 2021c) and against compulsory vaccinations in November 2021, calling Austria a "dictatorship" (APA OTS 2021d).

This clear policy cut – which stands in contrast to, e.g., the zig-zag course by Trump (Bisbee and Lee 2020) – makes the case of the Austrian FPÖ ideally suited to study three hypotheses regarding the "Downsian" and the "party identification" effect:

**Hypothesis I**: The electorate adapted their vote preferences after observing the policy turn ("weak Downsianism").

**Hypothesis IIa**: The FPÖ policy switch had an impact on the beliefs of voters who identify with the party about benefits and/or costs of containment measures ("party identification effect"). The alternative hypothesis would allude to a "strong Downsianism", in which the party stance does not affect voter preferences.

**Hypothesis IIb**: The FPÖ policy switch had an impact on the regional differentiation of the pandemic ("party identification effect"). Again, the alternative hypothesis would suggest "strong Downsianism".

In order to investigate these hypotheses empirically, I draw in this paper on both individual-level panel survey data, as well as administrative data on infections and deaths. While my analysis supports the first two hypotheses, the first test of hypothesis IIb delivers a puzzling result, namely that the policy turn seemed to have an impact on the regional distribution of the number of deaths per capita (producing a relative increase of the fatality rate in FPÖ strongholds), but not on the reported number of cases per capita. I develop a new hypothesis to solve this puzzle, namely that corona skeptics are less likely to test themselves for Covid-19 and hence the corona populist turn of the FPÖ also increased the dark figure (i.e. the share of undetected cases). In order to test and explore this hypothesis, I complement my analysis with individual-data from a prevalence study, as well as with simulations by using an extended version of the standard epidemiological model known as the SIR model which suggest that i) skepticism is indeed correlated with the dark figure, i.e. skeptics are less likely to get tested and hence do not only increase the true number of infections, but also the undetected share of infections, and ii) that such a self-selection bias is a plausible solution to the abovementioned puzzle within a formal theoretical framework.

With this paper, I hence aim to contribute to our understanding of i) the relation between the policy platform of political parties and the views and behavior of voters (e.g. Bechtel et al. 2015; Downs 1957; Druckman et al. 2013; Grewenig et al. 2020), ii) the role of politics in the Covid-19 pandemic (e.g. Allcott et al. 2020; Fan et al. 2020; Gollwitzer et al. 2020; Milosh et al. 2021), which is part of a broader research agenda on the causes and effects of behavioral differences in the pandemic (e.g. Bai and Brauer 2021; Barrios et al. 2020; Brzezinski et al. 2020; Bursztyn et al. 2020; Chernozhukov et al. 2020; Jung et al. 2020; Papageorge et al. 2020; Wright et al. 2020), iii) the implications of heterogeneous mixing in the context of epidemiological models, which are studied both with classical SIR (e.g. Acemoglu et al. 2021; Britton et al. 2020; Ellison 2020; Bursztyn et al. 2020) and agent-based approaches (e.g. Basurto et al. 2021; Delli Gatti and Reissl 2022; Mellacher 2020).

### *1.1 Related literature about the Covid-19 pandemic*

The effects of political polarization and populism on beliefs, behavior, and public health have attracted scholarly attention across scientific disciplines (Alashoor et al. 2020; Brubaker 2020; Eberl et al. 2021; Lasco 2020; McKee et al. 2020; Pevehouse 2020).

Most research has concentrated on the US: For instance, Allcott et al. (2020) show using mobile phone data on the county level that democratic counties exercise more social distancing (also confirmed by, e.g., Baradaran Motie and Biolsi 2021), but also record more cases and deaths per capita than republican counties. Controlling for a large number of covariates, Gollwitzer et al. (2020) however find that Trump-leaning counties do not only exercise less social distancing, but that this is also linked to higher growth rates in the number of cases and fatalities.

Allcott et al. (2020) also confirm that individual beliefs about the severity of Covid-19 are linked to self-reported social distancing using data from an online survey with US participants. Further investigating what drives these differences, Fan et al. (2020) document that there are partisan differences in social distancing behavior and beliefs, which also depend on differences in news consumption using data from an online survey. Wu and Huber (2021) use a regression analysis of survey data to show that partisan differences in self-reported social distancing disappear once they control for beliefs and social norms.

Bisbee and Lee (2020) show that Republican-leaning counties were more likely to practice social distancing when Trump emphasized the risks of Covid-19 on his Twitter profile. As seen in their analysis, however, Trump sent at best a mixed message about the severity of Covid-19, making causal analysis difficult.

Research on other countries than the US is sparser. Barbieri and Bonini (2021) show that a higher vote share for the Italian right-wing party Lega is associated with a lower degree of social distancing using regional mobility data. Like Trump's course, the Lega's policy was characterized by a zig-zag course: first downplaying the pandemic, then agreeing to a lockdown, followed by a call for a fast re-opening. In February 2021, the party entered into a "national unity" coalition government.

Eberl et al. (2021) show that "populist" attitudes – which they define as being anti-elitist, people-centered and having a "Manichean outlook" (following Hawkins and Rovira Kaltwasser 2018) – are positively correlated with Covid-19 conspiracy theories in Austria using data from the Austrian Corona Panel Project (Kittel et al. 2020, 2021). Eberl et al. (2021) emphasize, however, that such views are to be found everywhere in the left-right spectrum and not tied to voters of the FPÖ specifically. Charron et al. (2022) show that excess mortality is higher in European regions where elite polarization is stronger in the dimension of European integration, which they argue proxies the strength of populism.

The next section discusses the Austrian empirics. The third section is devoted to the extended SIRD model and its implications. The fourth section concludes.

## 2 Empirics

This section is split in three parts. In the first part (2.1), I use panel survey data from the Austrian Corona Panel Project (Kittel et al. 2020, 2021) to investigate the impact of the policy turn on a) the voting intentions of skeptics vs. non-skeptics among non-FPÖ voters (i.e. the "Downsian" effect), and b) the beliefs of FPÖ voters regarding the danger posed by Covid-19 vis-à-vis non-FPÖ voters (i.e. the "party identification" effect) using a difference-in-differences approach.

In the second part (2.2), I use aggregate-level administrative data on infections and deaths to investigate the impact of the turn on the regional differentiation of the pandemic in Austria.

Finally, I use data from the Austrian Covid-19 prevalence study conducted in November 2020 to study the reported and true infection status of "corona skeptics" and non-skeptics in subsection 2.3. This data source includes information about the true and reported infection status, as well as information about the reported policy stance towards government policy, hence allowing me to connect beliefs with epidemiological characteristics on an individual level.

*2.1.1 Individual-level panel survey evidence: Data & Method*

In this subsection, I use data from the Austrian Corona Panel Project (Kittel et al. 2020, 2021) to explore how the policy switch affected individual beliefs and voting intentions. This data source includes information about partisan affiliation, demographic factors and a wide array of questions on beliefs and perceptions, hence providing a comprehensive overview of the evolution of beliefs on the Covid-19 pandemic in Austria. While the panel is not balanced, i.e. not every respondent participated in every survey "wave", new respondents are filled in to match the demographic and political characteristics of those who have to be replaced. Every wave includes approximately 1500 survey respondents. In my analysis, I use data from waves 1-28, omitting wave 5, as it was conducted during the policy switch of the FPÖ on the 27th of April 2020.

Since fig. 1 suggests a mutually reinforcing relationship between populism and skepticism, my empirical strategy has to deal with an "endogeneity problem" – it is a priori unclear, whether people who declare that they want to vote for the FPÖ and are corona skeptics are FPÖ voters because of their skepticism, or whether they are skeptic because they identify with the policy platform of the FPÖ.

In order to differentiate between the "Downsian" effect (i.e. increased support for the FPÖ among skeptics and/or reduced support among others) and the "party identification effect" (i.e. increased skepticism among FPÖ partisans), I rely on a survey item that asks respondents which political party they voted for in 2019, i.e. the year before the Covid-19 crisis and hence eliminating any endogeneity concerns for this particular variable. Respondents who answered that they were enfranchised at that time, but did not vote for the FPÖ can reasonably be expected not to be exposed to a "party identification effect". Hence, I use this group to study the "Downsian" hypothesis and as a control group to study the "party identification effect", for which I assume that the treatment group is composed of those respondents who stated that they did vote for the FPÖ in 2019.

While the fact that an individual has voted for the FPÖ in 2019 cannot be influenced by the Covid-19 crisis starting in 2020, the interaction term between this variable and the policy turn, i.e., may still be compromised by endogeneity if the policy turn of the FPÖ does not represent a reaction to a general rise in skepticism within the whole population, but only within their voters. While the policy turn did not significantly affect voting intentions among people who voted for the FPÖ in 2019 and have embraced corona skepticism, it reduced the support for the FPÖ among their former voters who remained non-skeptical. This result alleviates concerns that the interaction term between voting for the FPÖ in 2019 and the policy turn is endogenous. Instead of solidifying the FPÖ's voter base, the policy turn seemed to have shifted it towards more skeptic parts of the electorate.

*2.1.2 Individual-level panel survey evidence: "Downsian" effect*

In this first analysis, I study whether voters reacted to the policy turn of the FPÖ by changing their voting preferences using data from the Austrian Corona Panel Project (Kittel et al. 2020, 2021). For this analysis, I rely on two survey items. The first asks respondents how likely they will ever vote for the FPÖ on a 11-point Likert scale from 0 (very unlikely) to 10 (very likely). Complete data on this item has only been collected in 10 out of the 28 total waves of the Austrian Corona Panel Project until January 2022 which are analyzed in this paper. The second item asks respondents whether they consider the response of the Austrian government to the Covid-19 crisis appropriate on a 5-point Likert scale from 1 (completely insufficient) to 5 (too extreme), whereas 3 means "appropriate". This data was collected in every wave. In order to facilitate the statistical analysis, I recode the two variables into i) a dummy variable "would likely vote for FPÖ" which is 1 for respondents who their likelihood to ever vote for the FPÖ between 6 and 10, and 0 otherwise, and ii) a dummy variable "measures are exaggerated", which is 1 for respondents who score 1 or 2 on the Likert scale capturing the individuals' stance on the appropriateness of containment measures and 0 otherwise, and finally iii) a dummy variable "measures are too lax" for those who score 4-5 on the same scale. Accordingly, respondents who score 0 for both of the two dummy variables "measures are exaggerated" and "measures are too lax" believe that the measures are appropriate.

Since voting intentions and policy preferences likely mutually influence themselves (see fig. 1), I conduct this exercise separately for respondents who stated that they voted for the FPÖ at the last national elections in 2019 (which were pre-Covid-19) and others. This approach is useful for at least two reasons: Respondents who did not state that they voted for the FPÖ in the pre-Covid-19 elections are much less likely to be exposed to a "party identification effect", hence allowing to better identify the "Downsian effect" of people who may change their voting preferences due to the turn, but likely not their policy preferences.

Figure 2 shows the mean and standard deviation of the dummy variable "would likely vote for the FPÖ" for those respondents who did not declare that they voted for the FPÖ in 2019. While the support for the FPÖ among survey respondents in this category decreased among all groups during the summer of 2020, we see a polarization starting in the autumn of 2020, where respondents who believe that the government measures are exaggerated are increasing their voting preferences for the FPÖ in relative and absolute terms. Likewise, we see an effect for the group of people who stated that they believe that the government measures are too lax, whose voting preferences for the FPÖ drastically dropped after the summer of 2020.

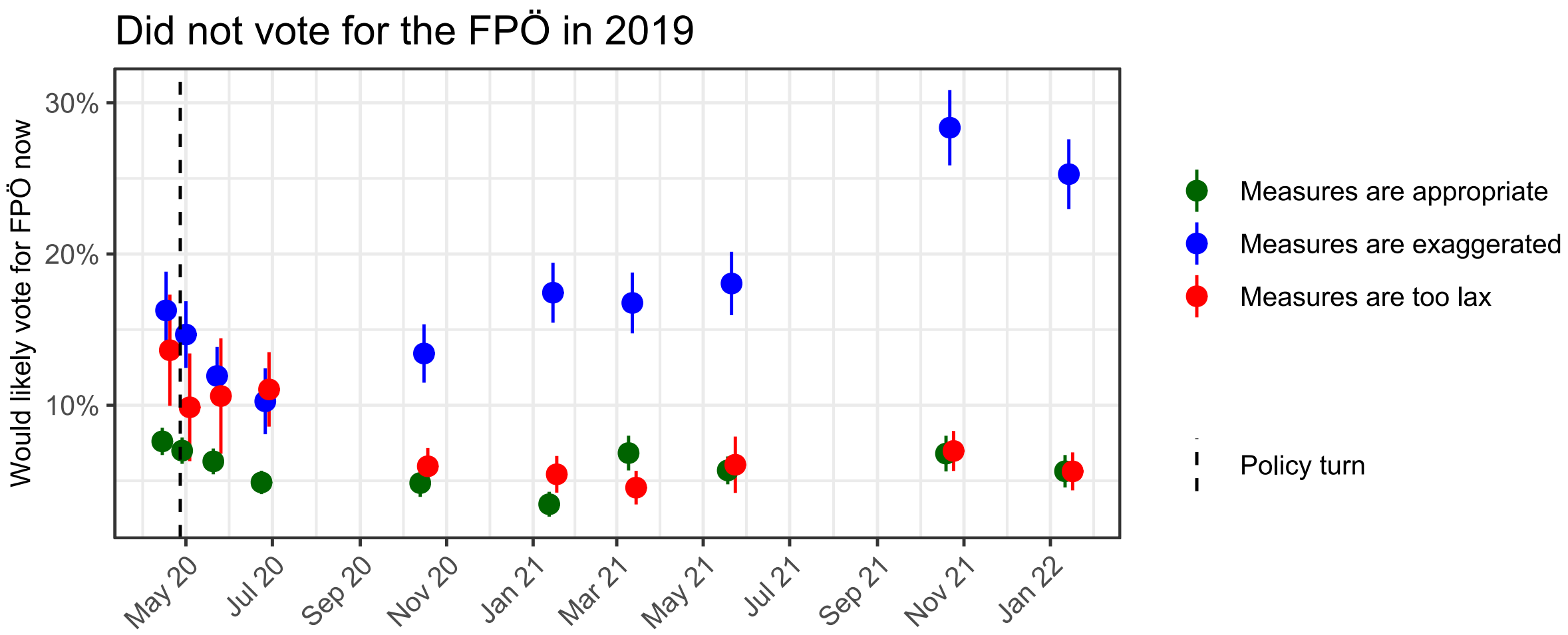

**Figure 2**: The evolution of voting preferences for the FPÖ of respondents who did not state that they voted for the FPÖ at the national elections of 2019

These results support the "weak Downsianism" hypothesis (Hypothesis I), i.e. that voters adapted their voting preferences according to their policy preferences. Interestingly, however, the trend between the groups only diverges starting in the autumn of 2020. A likely explanation for this fact is that the question of pandemic mitigation measures was less salient over the summer, as the number of infections was low and only a few restrictions remained in place.

Figure 3 shows the same set of results for respondents who declared that they voted for the FPÖ at the last national elections. The interpretation of these results is more complicated, as their policy position regarding the Covid-19 measures may likely also be influenced by the party's position (as I will investigate in the next subsection). Hence, we are facing an endogeneity problem that only allows us to note correlations and hypothesize about causations.

Nevertheless, this exercise is important. On the one hand, tracking the voting preferences of respondents who voted for the FPÖ previously is interesting in its own right. On the other hand, this exercise could shine some light on the question whether the policy U-turn has been driven by the desire to hold on to their traditional voter base. If this was true, the party's U-turn would merely be a reflection of a tendency that existed among (first and foremost) their voters. In this case, their voter base (or at least its corona skeptical part) would be discontent before the switch, and increase its support for the party after the switch.

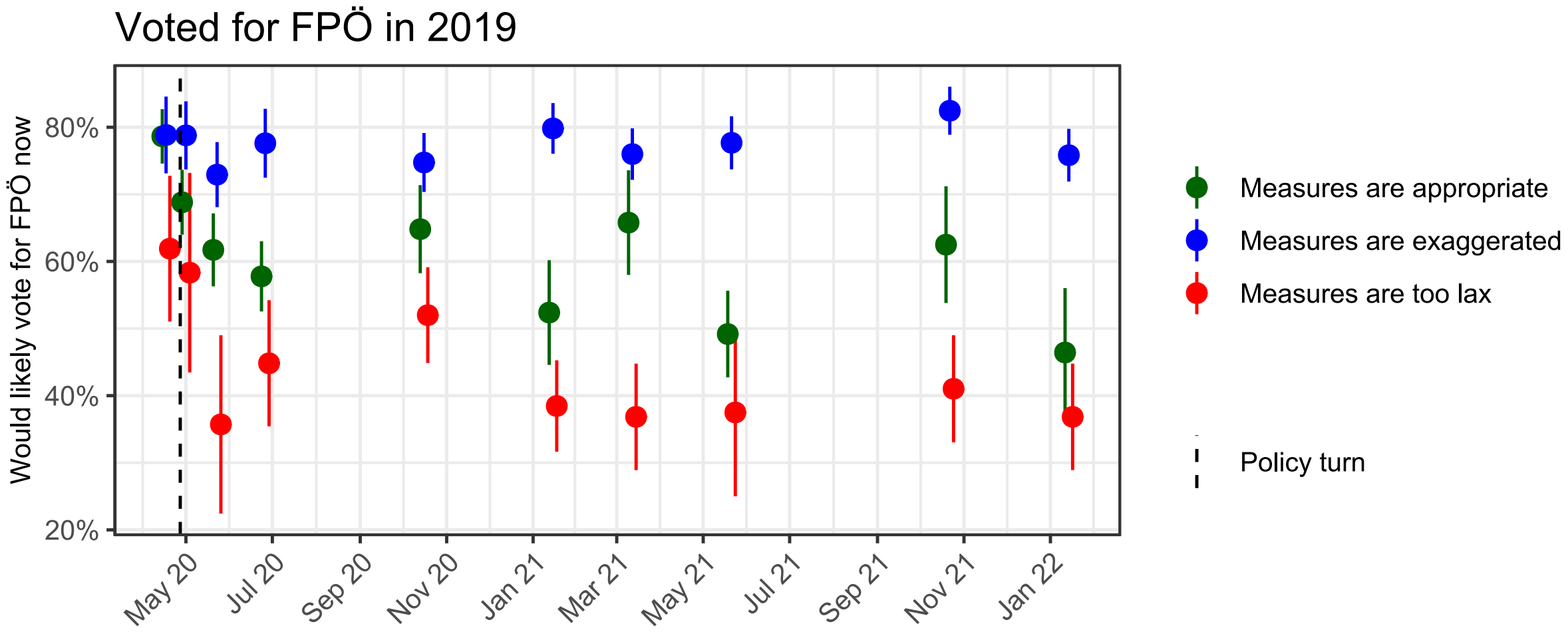

**Figure 3**: The evolution of voting preferences for the FPÖ of respondents who stated that they voted for the FPÖ at the national elections of 2019

Figure 3 indicates that this is not the case. Before the switch, those who believed that the measures are exaggerated were just as likely to state that they would vote for the party again as those who stated that the measures are appropriate. Immediately after the turn, this changed, however, as those who thought that the containment measures were appropriate were less likely to state that they would vote for the FPÖ again. It is interesting to see here that the turn seems to have triggered an immediate reaction among respondents who declared that they voted for the FPÖ in 2019 – in contrast to others, where the reaction seemed to have come only in autumn of 2020 (see fig. 2). However, due to endogeneity problems, it is unclear where this drop comes from: voters who were content with the government containment measures before the turn may have lowered their voting intentions for the FPÖ ("Downsian effect"), but they could also have adapted their policy stance to the one of their party ("party identification effect"). In this case, the drop may be a reflection of the fact that those who realigned their preferences following the policy turn were more content with the party to begin with.

In order to shed more light on the dynamics involved, figure 4 shows the evolution of voting intentions among the groups of a) those who declared that they voted for the FPÖ at the last national elections and b) others as a whole. After the turn, the share of respondents who declared that they would likely vote for the FPÖ dropped in both groups. However, while the support among people who did not declare that they voted for the FPÖ in 2019 recovered and eventually exceeded the value reached before the turn, the drop seemed to have been rather permanent in the group of FPÖ voters of 2019.

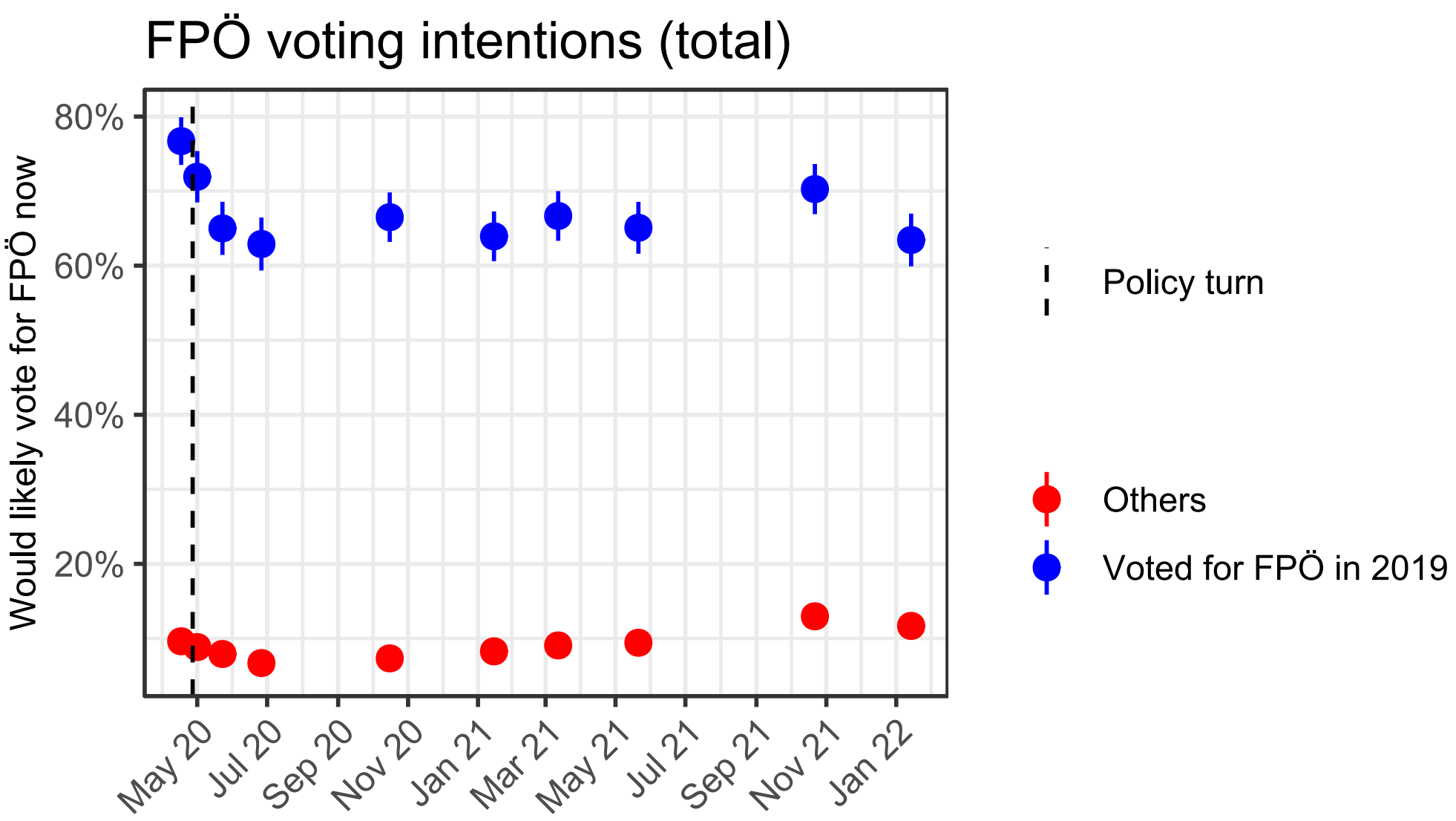

**Figure 4**: The evolution of voting preferences for the FPÖ of respondents who stated that they voted for the FPÖ at the national elections of 2019

In order to formally test these relationships, I estimate two types of models using probit regressions. The first specification is given by equation 1:

$$y_{i,t} = x_{i,t}^T\lambda + s_t\gamma + \left(s_t \times x_{i,t}^T\right)\rho + \alpha + \varepsilon_{i,t} \quad (1)$$

Where $y_{i,t}$ is the binary dependent variable, which is 1 for respondents who stated that they would likely ever vote for the FPÖ (between 6 and 10 on a Likert scale from 0-11) and 0 otherwise, $x_i^T$ is a vector of (dummy) variables capturing either the individual's policy stances (as defined above) or whether the individual declared that they voted for the FPÖ in 2019, $s_t$ a dummy variable for policy switch, which is 0 before the 27th of April and 1 afterwards, $\alpha$ the intercept and $\varepsilon_{i,t}$ is an error term.

In the second specification, shown in equation 2, I account for time fixed effects. In this case, the effect of the policy switch itself, as well as the intercept, are dummied out. Instead, this model includes time (=wave) fixed effects denoted by $\vartheta_t$:

$$y_{i,t} = x_{i,t}^T\lambda + \left(s_t \times x_{i,t}^T\right)\rho + \vartheta_t + \varepsilon_{i,t} \quad (2)$$

All regressions in this paper were conducted with the fixest package (Bergé 2018) for the programming language R (R Core Team 2021). The regression tables were prepared with the modelsummary package (Arel-Bundock 2022a). The standard errors are clustered for waves and respondents.

Table 1 shows the regression results. Models (1) – without time fixed effects – and (2) – with time FE – analyze how respondents who did not declare that they voted for the FPÖ in 2019 reacted to the policy switch. These results are most important in testing hypothesis I, as this group can plausibly be assumed not to be exposed to a FPÖ "party identification effect", hence minimizing the risk of endogeneity. Both models show that the support for the FPÖ among those who believe that the government response to Covid-19 is exaggerated increased after the switch, while it decreased among those who believe that it was too lax. In addition to that, model (1) shows that the support for the FPÖ among those who believe that the government measures are appropriate (i.e. the reference group) drops as well due to the switch. These results support hypothesis I, namely that the policy turn has influenced the voting preferences of the electorate ("weak Downsianism").

Models (3) and (4) show the same set of results, but only for those respondents who declared that they voted for the FPÖ at the elections of 2019, again with/without time fixed effects. In this group, the policy turn only significantly affected the voting intentions within the reference group (i.e. those who believe that the government measures were appropriate), as visible in model (3). It is important to note that the switch likely also has affected the *size* of this group, as shown in the next subsection.

Finally, models (5) and (6) test how the voting intentions of respondents who declared that they voted for the FPÖ in 2019 changed vis-à-vis others after the turn. These models are the only classical difference-in-differences specifications in this subsection, as the variable distinguishing the groups (i.e. voted for the FPÖ in 2019) is time-invariant. The models show that people who voted for the FPÖ in 2019 were less likely to vote for the FPÖ after the turn than before.

Table 1: Voting intentions for the FPÖ

| | **Group: Did not vote for FPÖ in 2019 only** | | **Group: Voted for FPÖ in 2019 only** | | **Group: All respondents** | |
|---|---|---|---|---|---|---|
| | **(1)** | **(2)** | **(3)** | **(4)** | **(5)** | **(6)** |
| Government response is exaggerated | 0.449*** | 0.449*** | 0.007 | 0.007 | | |
| | (0.025) | (0.027) | (0.055) | (0.060) | | |
| Government response is exaggerated x Policy switch | 0.208** | 0.196** | | | | |
| | (0.065) | (0.060) | | | | |
| Government response is too lax | 0.335*** | 0.335*** | -0.491*** | -0.491*** | | |
| | (0.011) | (0.009) | (0.065) | (0.073) | | |
| Government response is too lax x Policy switch | -0.283*** | -0.297** | 0.038 | 0.051 | | |
| | (0.059) | (0.064) | (0.043) | (0.046) | | |
| Voted for FPÖ in 2019 | | | | | 2.033*** | 2.033*** |
| | | | | | (0.043) | (0.043) |
| Voted for FPÖ in 2019 x Policy switch | | | | | -0.283*** | -0.275*** |
| | | | | | (0.034) | (0.034) |
| Policy switch | -0.142*** | | -0.543*** | | -0.030 | |
| | (0.024) | | (0.060) | | (0.034) | |
| Intercept | -1.432*** | | 0.794*** | | -1.304*** | |
| | (0.020) | | (0.040) | | (0.017) | |
| Num.Obs. | 11747 | 11747 | 1885 | 1885 | 13632 | 13632 |
| R2 | 0.046 | 0.050 | | 0.064 | 0.234 | 0.238 |
| R2 Adj. | 0.045 | 0.047 | | 0.053 | 0.234 | 0.236 |

Table 1: Voting intentions for the FPÖ

| | **Group: Did not vote for FPÖ in 2019 only** | | **Group: Voted for FPÖ in 2019 only** | | **Group: All respondents** | |
|---|---|---|---|---|---|---|
| | **(1)** | **(2)** | **(3)** | **(4)** | **(5)** | **(6)** |
| R2 Within | | 0.044 | | 0.058 | | 0.236 |
| R2 Within Adj. | | 0.043 | | 0.055 | | 0.236 |
| AIC | | | 2255.6 | | | |
| BIC | | | 2288.9 | | | |
| Log.Lik. | | | -1121.811 | | | |
| RMSE | 0.28 | 0.28 | 0.45 | 0.45 | 0.32 | 0.32 |
| Std.Errors | by: wave & respid | by: wave & respid | by: wave & respid | by: wave & respid | by: wave & respid | by: wave & respid |
| FE: wave | | X | | X | | X |

* p < 0.05, ** p < 0.01, *** p < 0.001

Together, these results indicate that the policy U-turn i) was perceived by voters, who adapted their preferences accordingly (supporting hypothesis I: "weak Downsianism"), ii) shifted the voter base of the FPÖ, as it cost them some support among their voters in 2019, but gained them support among other parts of the population which embraced corona skepticism. The latter finding is also important as it alleviates concerns that the policy turn was mostly driven by FPÖ partisans themselves, and hence allow us to study the impact of the policy turn on FPÖ partisans with more credibility in the next subsection.

### *2.1.3 Individual-level panel survey evidence: Party identification hypothesis*

In order to study the effect of the FPÖ policy turn on the beliefs and policy preferences of FPÖ partisans vs. others, I again use data from the Austrian Corona Panel Project and focus on five survey questions that are initially coded as five-point Likert scales. Apart from the survey item about the appropriateness of current government containment policies already used in the previous analysis, I additionally use survey questions about the perceptions about the danger that Covid-19 poses to the private and public health, the private economic situation and the economy as a whole.

These items are included in every wave analyzed in this paper. Again, I recoded the Likert scales to create 10 dummy variables which are 1 for initial values for 1,2 (such as low public health danger) or 4,5 (such as high danger to personal economic situation). This recoding exercise helps to a) account for potential different behavior at the lower and upper end of the distribution, and b) facilitate the statistical analysis with the help of probit regressions.

Figure 5 shows the evolution of beliefs of declared FPÖ voters (respondents who declared that they voted for the FPÖ in the last national elections) vs. non-FPÖ voters (all other respondents) according to this data, omitting one wave that was conducted during the FPÖ policy switch.

A quick graphical analysis suggests that health perceptions were closely aligned between the two groups before the switch, but diverged after the policy switch in the sense that FPÖ voters were more likely to believe that Covid-19 poses i) a low danger to public health and ii) own personal health, and iii) less likely to believe in a high danger to public health.

On the other hand, there are pre-existing differences between the two groups with regard to their beliefs about the economic impact of the crisis and their policy views. FPÖ voters seemed to have been more likely to believe in a stronger impact on the personal economic situation (and less likely in a weaker impact) even before the policy switch and these differences persist after the switch. Curiously, FPÖ voters have been more likely to oppose government policy on Covid-19 in both directions, i.e. they were more likely to believe that government action against Covid-19 is exaggerated *and* that it was too lax than other survey respondents. However, the policy switch seemed to have had a coordinating effect in this matter, as the support among FPÖ voters for the view that the measures are exaggerated increased drastically after the switch, while the support for the opposing view diminished to a point where FPÖ voters were less likely to hold that view than the rest of the population.

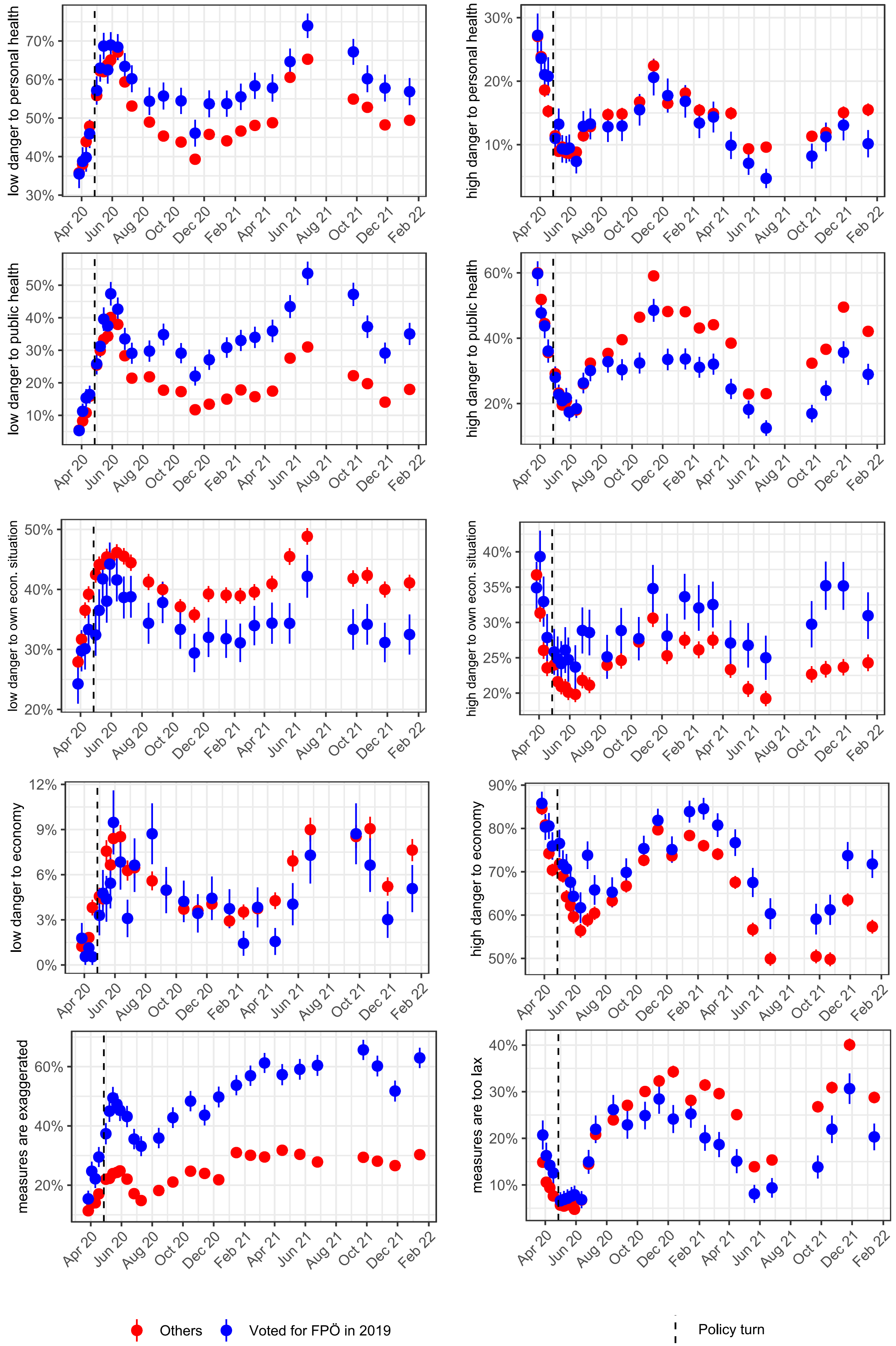

**Figure 5**: The evolution of beliefs of respondents who that they voted for the FPÖ at the national elections of 2019 vs. other respondents

I then proceed to an econometric analysis based on a difference-in-differences approach using probit regressions and controlling for potentially confounding factors such as age and gender. More precisely, I estimate the following model:

$$y_{i,t} = x_i^T\theta + f_i\lambda + s_t\gamma + (s_t \times f_i)\rho + (s_t \times x_i^T)\delta + \vartheta_t + \varepsilon_{i,t} \quad (3)$$

Where $y_{i,t}$ is the respective (binary) dependent variable, $x_i^T$ is a vector of (dummy) control variables, $f_i$ a dummy variable which is 0 for survey participants who declared to have voted for the FPÖ at the last national elections, $s_t$ a dummy variable for policy switch, which is 0 before the 27th of April (i.e. waves 1-4) and 1 afterwards, $z_t^T$ is another vector of (dummy) control variables, $\vartheta_t$ are time (=wave) fixed effects and $\varepsilon_{i,t}$ is an error term. I am particularly interested in the coefficient $\rho$, as it aims to evaluate whether the policy switch had an effect on the beliefs of FPÖ voters vis-à-vis non-FPÖ voters in a particular dimension. In order to account for a potential differential impact of socio-demographic variables before and after the policy switch (i.e. potential endogeneity, because the demographic characteristics are themselves correlated with support for the FPÖ), I also included an interaction term between the vector of control variables and the policy switch.

Table 2 shows the first series of regression results regarding beliefs about government policy using different models. The full regression tables involving the coefficients for all control variables can be found in appendix A (table A1). The standard errors are clustered for waves and respondents. The regressions were computed with the unweighted dataset, but appendix B shows regressions using demographic (tables B1-3) and political (tables B4-6) weights supplied by the ACPP, which are very similar to the unweighted results.

Simple and more sophisticated specifications agree with the graphical intuition on the following main insights: First, before the policy switch, declared FPÖ voters were both more likely to believe that government action against Covid-19 is exaggerated *and* that government action is too lax. Hence, beliefs about whether government action is appropriate were more polarized than those of respondents who were not declared FPÖ voters. Second, the switch seemed to have put an end to the disagreement among FPÖ-voters about *why* government policy should be opposed: On the one hand, the difference in the share of FPÖ voters vs. non-FPÖ voters who believed that government response against Covid-19 is exaggerated increased significantly after the switch. On the other hand, the support for the view that government policy is too lax after the switch among FPÖ voters declined up to a point where they were less likely to hold that view than the rest of the population.[1] Hence, the policy U-turn reduced the level of polarization regarding the appropriate level of containment policies among FPÖ voters.

Models (1) and (6) are specified in a very simple way by excluding any additional control variables. Models (2) and (7) also include gender-, age- and education-specific dummy variables. The full regression tables, to be found in appendix A (table A1), show several statistically significant correlations between the dependent variables and the control variables:

First, men were more likely to think that the containment strategy is exaggerated, but that the difference between men and women became smaller after the policy switch. This effect could be driven by gender differences in risk perceptions (Gustafsod 1998) and in the willingness to take risks (Charness and Gneezy 2012).

Second, young people below the age of 25, as well as (only after the switch) people above the age of 64 were less likely to think that government policy is exaggerated than the reference group (i.e. those between 25 and 64). Furthermore, people above 64 were less likely to think that the government response is too lax before the switch, but this age-specific effect reversed after the switch. This result may be driven by the different nature of the infection waves in Austria. As I will show in the subsection 2.2, only very few people died during the first wave of infections (i.e. before the policy switch), which was rather well-contained whereas the second wave in autumn of 2020 was particularly deadly. Hence, the experience with a less-contained spread of the virus may have caused men and people who are at risk due to their age to realize the necessity for (some) virus containment policies. The result that the youth has been less critical about government policy is at first glance counterintuitive, as they are at least risk from Covid-19. However, previous research from Jordan et al. (2021) has shown that pro-social motives are more powerful in encouraging social distancing behavior than personal ones. Hence, young people may be more inclined to comply with social distancing, as they are convinced that they are staying at home for others.

---

[1] Due to the opposite signs of the coefficients, I computed the conditional marginal effect with the marginaleffects package (Arel-Bundock 2022b) for R in order to verify this statement.

Finally, education matters: University graduates were overall less critical towards government policy than the reference group (i.e. people who finished at maximum compulsory education), although the share of university graduates who believed that the government response is too lax increased significantly after the policy switch relative to the reference group. Furthermore, people who completed at most an apprenticeship were less likely to believe that government policy is exaggerated before the switch, but more likely to do so after the switch. Education is a proxy for two different factors: First, their type of job may influence their exposure to the virus and the crisis both with regard to health and the economy (i.e. more or less "rational" factors which may be counteractive). Second, a higher level of education was shown by Eberl et al. (2021) to decrease the level of science skepticism and the belief in conspiracy theories (i.e. "irrational" factors).

Finally, models (3), (4), (7) and (8) incorporate the beliefs about the danger that Covid-19 poses to personal and public health, as well as to the own economic situation and the economy as a whole. In regressions (3) and (7), the reference group does not believe that Covid-19 poses a low danger to the economy/health etc., whereas the reference group in regressions (4) and (8) does not believe that it poses a high danger.

Expectedly, respondents who believe that Covid-19 poses a high risk to personal or public health are more likely to believe that government policy is too lax. Conversely, beliefs in low health risks are associated with the belief that government response is exaggerated. The opposite relationship can be observed for respondents who believe that Covid-19 poses a high danger to their personal economic situation or to the economy as a whole, as they were more likely to think that the government response was exaggerated and, in the case of the belief in a high danger to the economy, less likely to think that it was too lax. The opposite is true for respondents who believe that Covid-19 poses a low danger to their personal economic situation.

The result that the demand for containment policies is driven by beliefs regarding the danger of the Covid-19 crisis to health and economy offers support to the work by Allcott et al. (2020) and Proaño and Makarewicz (2021). In these models, social distancing and skepticism essentially represents a trade-off between health and economy, even though some other authors argue that this trade-off may not exist on a societal level (see, e.g., Bethune and Korinek 2020, who argue that an eradication strategy is superior in both dimensions).

Table 2: Opinion on government policy

| | Government response is exaggerated | | | | Government response is too lax | | | |
|---|---|---|---|---|---|---|---|---|
| | (1) | (2) | (3) | (4) | (5) | (6) | (7) | (8) |
| Voted for FPÖ in 2019 | 0.338*** | 0.299*** | 0.287*** | 0.285*** | 0.254*** | 0.228** | 0.250*** | 0.272*** |
| | (0.076) | (0.074) | (0.072) | (0.076) | (0.072) | (0.076) | (0.075) | (0.077) |
| Voted for FPÖ in 2019 x Policy switch | 0.339*** | 0.350*** | 0.289*** | 0.319*** | -0.415*** | -0.378*** | -0.312*** | -0.357*** |
| | (0.070) | (0.069) | (0.070) | (0.070) | (0.081) | (0.083) | (0.079) | (0.081) |
| Gender | | X | X | X | | X | X | X |
| Age | | X | X | X | | X | X | X |
| Education | | X | X | X | | X | X | X |
| Low danger to health/economy | | | X | | | | X | |
| High danger to health/economy | | | | X | | | | X |
| Wave FE | X | X | X | X | X | X | X | X |
| Num.Obs. | 41056 | 41056 | 41056 | 41056 | 41056 | 41056 | 41056 | 41056 |
| Pseudo R2 | 0.046 | 0.053 | 0.211 | 0.171 | 0.075 | 0.078 | 0.131 | 0.156 |
| Pseudo R2 Adj. | 0.044 | 0.051 | 0.209 | 0.169 | 0.074 | 0.076 | 0.129 | 0.154 |
| Pseudo R2 Within | 0.025 | 0.032 | 0.194 | 0.153 | 0.002 | 0.004 | 0.062 | 0.089 |
| Pseudo R2 Within Adj. | 0.025 | 0.032 | 0.193 | 0.152 | 0.002 | 0.004 | 0.061 | 0.088 |
| AIC | 45161.1 | 44826.4 | 37367.8 | 39273.1 | 37290.1 | 37221.2 | 35069.7 | 34077.5 |
| BIC | 45178.4 | 44947.1 | 37523.1 | 39428.3 | 37307.3 | 37341.9 | 35224.9 | 34232.8 |
| RMSE | 0.43 | 0.43 | 0.38 | 0.39 | 0.38 | 0.38 | 0.37 | 0.36 |
| Std.Errors | by: wave & respid | by: wave & respid | by: wave & respid | by: wave & respid | by: wave & respid | by: wave & respid | by: wave & respid | by: wave & respid |

* $p < 0.05$, ** $p < 0.01$, *** $p < 0.001$

In table 3, I show the results of probit regressions with different dependent variables, namely beliefs connected to a low danger to health and economy. The simple model (1), as well as the model involving socio-demographic control variables (2) agree that FPÖ voters did not differ significantly from non-FPÖ voters with regard to the belief that Covid-19 poses a low danger to personal health before the policy switch. After the switch, however, FPÖ voters were more likely to adhere to this belief.

Both the simple model (3) and the model (4) controlling for socio-demographic characteristics agree that FPÖ voters did not significantly differ from others in their belief that Covid-19 poses little danger to public health. Again, FPÖ voters are more likely to hold that belief after the switch in both models.

Models (5) and (6) show that FPÖ voters have been less likely to believe that Covid-19 poses little danger to their personal economic situation even before the switch and suggest that the policy switch did not have an impact in this regard. Finally, models (7) and (8) suggest that FPÖ voters did not differ with regard to the belief that Covid poses a low danger to the economy as a whole before or after the switch.

Again, the full regression tables are depicted in appendix A (table A2) and show the following significant correlations: Men are more likely to believe that Covid poses a low danger in all dimensions, whereas young people are more likely to believe in a low personal health danger than the reference group (i.e. people aged 25-64). Curiously, respondents above the age of 64 are more likely to believe that Covid-19 poses little health danger to them personally after the switch. This counterintuitive effect may stem from three causes: 1.) this group is usually retired and hence arguable more able to practice social distancing, 2.) it had sooner access to vaccines[2], 3.) potential psychological reasons such as denial. People above the retirement age are also more likely to believe that Covid hat little impact on their personal economic situation, most probably because almost all of them are retired and believe that the crisis would not have an impact on their pensions.

Finally, university graduates were more likely to believe in a low danger to personal health than people who finished at maximum compulsory education before the switch, but the differences disappeared afterwards. The initial effect may be driven by the differing teleworking capabilities.

[2] A more detailed analysis shows that people aged above 64 believed in low personal risks in the summer of 2020, i.e. when the incidence was low, as well as after every person in this group had the chance to become vaccinated (i.e. from the summer of 2021), but not during the deadly autumn and winter of 2020.

Table 3: Low danger to health / economy

| | Covid poses low danger to personal health | | Covid poses low danger to public health | | Covid poses low danger to personal economic situation | | Covid poses low danger to economy | |
|---|---|---|---|---|---|---|---|---|
| | **(1)** | **(2)** | **(3)** | **(4)** | **(5)** | **(6)** | **(7)** | **(8)** |
| Voted for FPÖ in 2019 | -0.036 | 0.041 | 0.107 | 0.082 | -0.126* | -0.114 | -0.285 | -0.279 |
| | (0.067) | (0.071) | (0.084) | (0.082) | (0.064) | (0.064) | (0.178) | (0.187) |
| Voted for FPÖ in 2019 x Policy switch | 0.209*** | 0.137* | 0.266** | 0.277*** | -0.046 | -0.067 | 0.205 | 0.192 |
| | (0.054) | (0.057) | (0.083) | (0.081) | (0.043) | (0.046) | (0.183) | (0.192) |
| Gender | | X | | X | | X | | X |
| Age | | X | | X | | X | | X |
| Education | | X | | X | | X | | X |
| Wave FE | X | X | X | X | X | X | X | X |
| Num.Obs. | 41056 | 41056 | 41056 | 41056 | 41056 | 41056 | 41056 | 41056 |
| Pseudo R2 | 0.023 | 0.032 | 0.050 | 0.062 | 0.007 | 0.025 | 0.027 | 0.041 |
| Pseudo R2 Adj. | 0.023 | 0.031 | 0.049 | 0.060 | 0.006 | 0.024 | 0.024 | 0.037 |
| Pseudo R2 Within | 0.001 | 0.011 | 0.008 | 0.020 | 0.001 | 0.019 | 0.001 | 0.016 |
| Pseudo R2 Within Adj. | 0.001 | 0.010 | 0.008 | 0.019 | 0.001 | 0.019 | 0.001 | 0.014 |
| AIC | 55498.7 | 55012.9 | 41522.1 | 41028.5 | 54862.8 | 53919.4 | 16275.9 | 16066.0 |
| BIC | 55515.9 | 55133.6 | 41539.4 | 41149.2 | 54880.0 | 54040.1 | 16293.1 | 16186.7 |
| RMSE | 0.49 | 0.49 | 0.41 | 0.40 | 0.49 | 0.48 | 0.22 | 0.22 |
| Std.Errors | by: wave & respid | by: wave & respid | by: wave & respid | by: wave & respid | by: wave & respid | by: wave & respid | by: wave & respid | by: wave & respid |

* p < 0.05, ** p < 0.01, *** p < 0.001

Finally, table 4 computes regressions regarding the beliefs that Covid-19 poses a high danger to health or economy. Again, FPÖ voters did not differ from others regarding their health perceptions before the switch. While only the simple model (1) suggests that the policy switch caused FPÖ voters to believe less likely in a high danger to *personal* health, both models (3) and (4) suggest that this is the case for the high danger to *public* health. While FPÖ voters did not significantly differ from others with regard to their beliefs in a high danger to their personal economic situation (see models 5 and 6), they have been more likely to believe that Covid poses a high danger to the economy as a whole throughout the observation period (see models 7 and 8). The detailed results regarding gender, age and education are again shown in appendix A (table A3) and mostly mirror the results from table A2.

Table 4: High danger to health/economy

| | Covid poses high danger to personal health | | Covid poses high danger to public health | | Covid poses high danger to personal economic situation | | Covid poses high danger to economy | |
|---|---|---|---|---|---|---|---|---|
| | (1) | (2) | (3) | (4) | (5) | (6) | (7) | (8) |
| Voted for FPÖ in 2019 | 0.070 | -0.024 | -0.029 | -0.017 | 0.126 | 0.074 | 0.118 | 0.124 |
| | (0.077) | (0.082) | (0.062) | (0.062) | (0.078) | (0.079) | (0.067) | (0.067) |
| Voted for FPÖ in 2019 x Policy switch | -0.129 | -0.054 | -0.197** | -0.199** | 0.034 | 0.083 | 0.052 | 0.058 |
| | (0.068) | (0.072) | (0.062) | (0.062) | (0.067) | (0.070) | (0.060) | (0.060) |
| Gender | | X | | X | | X | | X |
| Age | | X | | X | | X | | X |
| Education | | X | | X | | X | | X |
| Wave FE | X | X | X | X | X | X | X | X |
| Num.Obs. | 41056 | 41056 | 41056 | 41056 | 41056 | 41056 | 41056 | 41056 |
| Pseudo R2 | 0.021 | 0.032 | 0.050 | 0.058 | 0.008 | 0.022 | 0.030 | 0.039 |
| Pseudo R2 Adj. | 0.020 | 0.030 | 0.049 | 0.056 | 0.007 | 0.021 | 0.029 | 0.037 |
| Pseudo R2 Within | 0.000 | 0.012 | 0.002 | 0.011 | 0.001 | 0.016 | 0.001 | 0.010 |
| Pseudo R2 Within Adj. | 0.000 | 0.011 | 0.002 | 0.010 | 0.001 | 0.015 | 0.001 | 0.010 |
| AIC | 32958.7 | 32606.3 | 50937.1 | 50522.2 | 45873.4 | 45239.1 | 50366.8 | 49950.3 |
| BIC | 32975.9 | 32727.0 | 50954.3 | 50642.9 | 45890.6 | 45359.8 | 50384.0 | 50071.0 |
| RMSE | 0.35 | 0.34 | 0.46 | 0.46 | 0.43 | 0.43 | 0.46 | 0.46 |
| Std.Errors | by: wave & respid | by: wave & respid | by: wave & respid | by: wave & respid | by: wave & respid | by: wave & respid | by: wave & respid | by: wave & respid |
| FE: wave | X | X | X | X | X | X | X | X |

* $p < 0.05$, ** $p < 0.01$, *** $p < 0.001$

### *2.1.2 Individual-level panel survey evidence: robustness checks*

Appendix C explores whether these results change if we introduce declared voting intentions for other parties or being a declared non- or invalid voter as control variables. This is helpful for two reasons: First, by including another party, the reference group against which the FPÖ voters are compared changes. For instance, if we include ÖVP (the major conservative governmental party) voter as an additional control variable, the FPÖ voters are compared with survey respondents who are neither declared FPÖ voters, nor declared ÖVP voters. Second, this analysis can help to uncover whether the results for the FPÖ are mirrored by another party.

My main results are robust in each model. My analysis suggests that the policy switch:

1.) Increased the likelihood that FPÖ voters believed that government response against Covid-19 is exaggerated. This is not true for any other party.
2.) Reduced the likelihood that FPÖ voters believed that government response against Covid-19 is too lax. This is not true for any other party.
3.) Increased the likelihood that FPÖ voters believed that Covid-19 poses a low danger to their personal health after the policy switch. This is also true for SPÖ (the social democratic party) voters, although it is important to note that SPÖ voters are the only political group which was less likely to believe in a low danger to their personal health before the policy switch. Thus, the reduction for SPÖ voters represented a move to the mean belief, whereas the reduction for FPÖ voters represented a move away from the mean belief.
4.) Increased the likelihood that FPÖ voters believed that Covid-19 poses a low danger to public health after the policy switch. This is not true for any other party.
5.) Reduced the likelihood that FPÖ voters believed that Covid-19 poses a high danger to public health after the policy switch. This is also true for SPÖ voters – but again, SPÖ voters are the only political group that had been significantly more likely to hold this belief before the switch. Hence, this effect could also be driven by a regression to the mean belief.

*2.1.3 Individual-level panel survey evidence: Summary*

My difference-in-differences analysis of the panel survey data from the Austrian Corona Panel Project (ACPP) (Kittel et al. 2020, 2021) suggests that the FPÖ policy switch affected the beliefs of people who voted for the FPÖ in the last pre-Covid-19 elections vis-à-vis others with regard to i) health perceptions, and ii) policy views, but did not significantly affect their perceptions regarding the economic consequences of the crisis which also play a role in explaining policy views, since people who are worried about the economic fallout of Covid-19 are less likely to support containment measures.

*2.2.1 Aggregate-level evidence: Data & Method*

In my second study, I investigate whether the policy stance of the FPÖ had an effect on the regional differentiation of the evolution of the pandemic in Austria. In order to do so, I draw on district-level data on the number of infections and deaths, which are available for a daily basis (BMSGPK 2022a). Studying county-level data is a standard approach followed by e.g. Allcott et al. (2020), Fan et al. (2020) and Gollwitzer et al. (2020) to study the impact of polarization on the spread of the virus in the US and districts are the Austrian counterpart for counties.

To get a first graphical intuition of the evolution of the pandemic in communities with a low or a high FPÖ vote share, I split the time series dataset into two groups, one for districts with a FPÖ vote share below or equal to and one above the median share of this party. Figure 6 shows the mean and standard deviation of the cumulated number of infections per 1,000 inhabitants and deaths per 100,000 inhabitants over time. This exercise suggests that districts in which the FPÖ fared relatively well at the last national elections received slightly lower damage in the first wave of infections, reporting lower numbers of cases and deaths, although the difference may not be statistically significant.

In the second infection wave starting in autumn 2020, however, the cumulated death toll in these districts surpasses the total number of deaths in the other districts, indicating that the second wave hit districts with a high FPÖ vote share much harder. The differences between the two groups then seem to be stable until the end of 2021, when the "Omicron" variant arrived in Austria. However, we do not observe the same clear trend in the cumulative number of reported cases per capita, as districts with a low FPÖ vote share continued to have a higher number of cumulative cases until the beginning of 2021, when there was already a clear difference with regard to mortality (see fig. 6). Even afterwards, districts with a high FPÖ vote share only had a small and potentially not statistically significant edge in terms of infections.

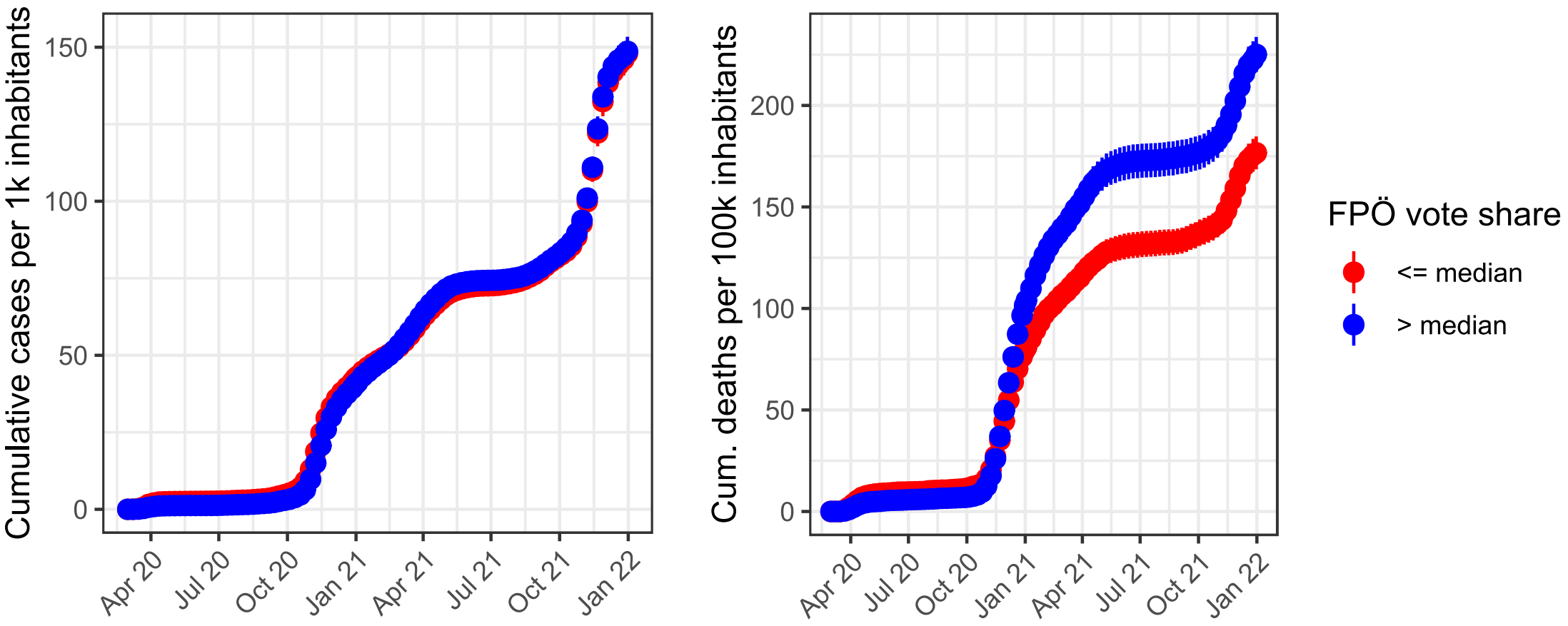

**Figure 6**: Cumulative cases per 1k inhabitants (left) and cumulative deaths per 100k inhabitants (right).

While the result regarding the number of deaths would indeed suggest that the FPÖ U-turn had an impact on the regional differentiation of the pandemic, the result regarding the number of cases is counterintuitive: after all, nobody can die from Covid-19 without contracting it. I develop an explanation for this phenomenon subsequently.

In order to confirm whether this graphical intuition is also statistically significant, especially when considering district-specific characteristics which may drive this pattern such as, e.g., the age structure of the population, I then turned to panel regression analysis.

In order to do so, I first create a balanced panel data set of weekly data on infections and deaths based on the daily data. The FPÖ campaign against the containment measures started on the 27th of April 2020, i.e. calendar week 18. I thus created a dummy variable which is 1 beginning with the 18th calendar week of 2020 and 0 before.[3] The data analyzed in this paper range until 31st of December 2021, which corresponds to the end of the infection wave surrounding the "Delta" variant, which was followed by the "Omicron" variant in 2022.

I then merged the dataset on infections and deaths with data on the results of the last national elections (BMI 2019) on the district level as a proxy for the influence of the FPÖ as well as regional data about vaccinations (BMSGPK 2022b, 2022c). Finally, my analysis in the preceding subsection, which utilizes data from the Austrian Corona Panel Project (ACPP) (Kittel et al. 2020, 2021) shows that FPÖ voters a) have differed from non-FPÖ voters before and after the switch with regard to perceptions about the danger to the economy and personal economic situation, and b) that some beliefs regarding health and economy changed after the policy switch for some socio-demographic characteristics.

[3] It is reasonable to assume that the campaign did not immediately translate into an increase, as we have to take into account, e.g., the time from infection to the start of the symptomatic phase, as well as a potential delay due to testing. However, my results are insensitive to reasonable adjustments in the timing of this dummy variable.

This result suggests that a two-way fixed effects estimator may be prone to an endogeneity problem, as some of these characteristics are correlated with the support for the FPÖ (such as, e.g., the level of education). In order to address this issue, I imputed time and district-specific beliefs by linking the individual-level survey data to the aggregate-level administrative data according to three observable characteristics: gender, age, education (Statistik Austria 2020a, 2020b; WKO 2020). This exercise aims to capture the district-time specific variation with regard to beliefs and perceptions that cannot be captured by the fixed effects. It is important to note that this exercise will inevitably also account for some of the impact of corona populism and hence tend to underestimate the effect of the FPÖ policy switch exactly due to the correlation of the FPÖ vote share with these characteristics. Nevertheless, this exercise is important to help to establish a “lower bound” of the impact of the FPÖ policy switch.

In order to impute the regional time-specific belief indices, I first split the survey respondents into eight different demographic groups (see fig. 7) and compute mean values for the perceptions in each of the 10 dimensions used in the analysis in the preceding subsection for each group and wave. I chose these criteria based on a) their relevance as explanatory factors of beliefs and perceptions as shown in the previous subsection, and b) data quality, as i) district-specific Austrian data does not allow us to fully disentangle age, gender and highest education and ii) young people likely did not complete their education yet, which would distort the meaning of, e.g., the category “highest education: compulsory schooling”.

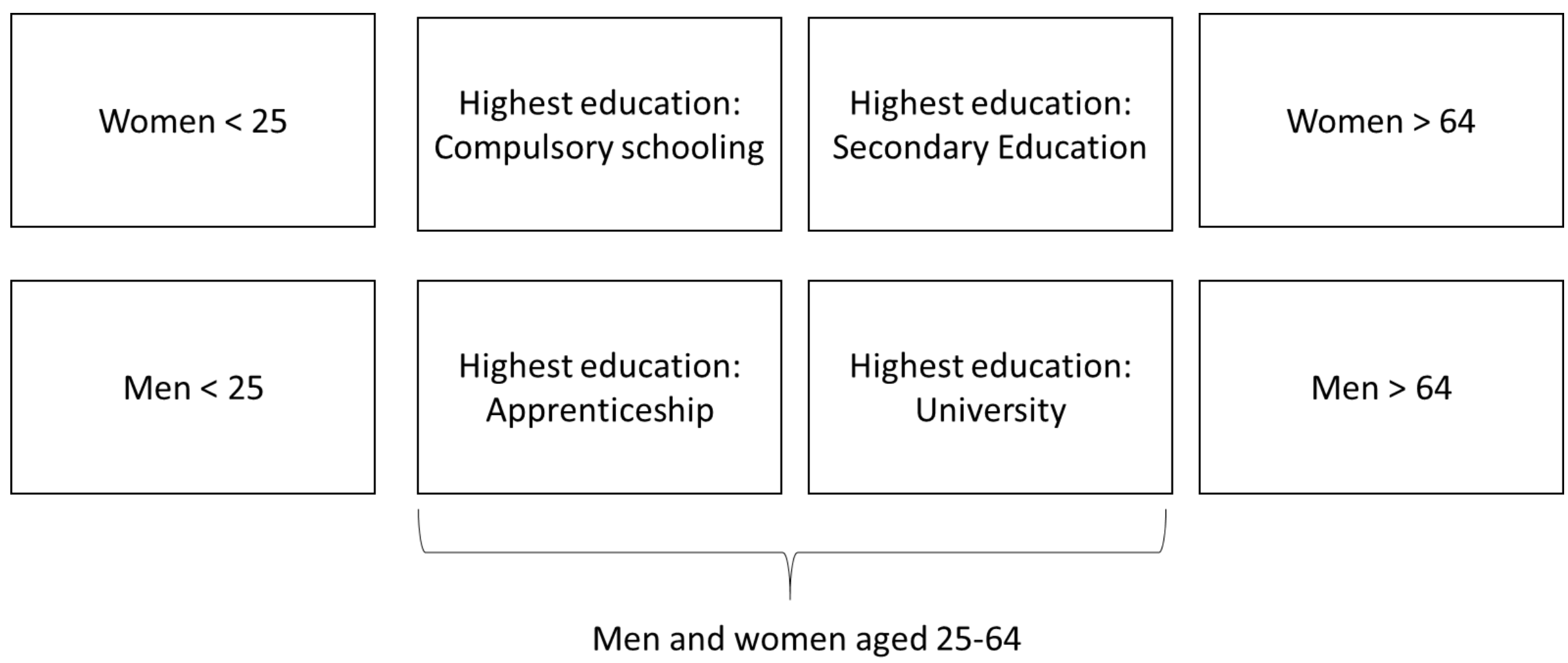

**Figure 7**: Demographic groups that are used to link data from the Austrian Corona Panel Project (Kittel et al. 2020, 2021) to the district-level dataset

If the mean belief for group $h$ with regard to dimension $d$ in week $t$ is $b_{d,h,t}$ and the share of group $h$ in district $i$ is $n_{i,h}$, then the imputed belief with regard to dimension d $b_{d,i,t}^{imp}$ is determined according to equation 4[4]:

$$b_{d,i,t}^{imp} = \sum_{h} n_{i,h} b_{d,h,t} \tag{4}$$

Using the combined dataset, I then estimate two-way fixed effects models of the following type:

$$y_{i,t} = y_{i,t-1}\alpha + x_{i,t}^{T}\vartheta + (s_t \times v_i)\rho + \theta_i + \eta_t + \varepsilon_{i,t} \tag{5}$$

Where $y_{i,t}$ is the dependent variable, either the number of deaths per 100k inhabitants or the number of reported cases per 1k inhabitants, $x_{i,t}$ a vector of control variables, $s_t$ a dummy indicating the policy switch, i.e. 1 after the FPÖ policy switch and 0 before, $v_i$ the FPÖ vote share in the specific district. District fixed effects are denoted with $\theta_i$ and time fixed effects with $\eta_t$. $\alpha$, $\gamma$, $\varphi$ and $\rho$ are coefficients, and $\vartheta$ is a vector of coefficients. Finally, $\varepsilon_{i,t}$ is the error term.

[4] Since the data on infections and deaths is weekly, but the surveys from the ACPP were not collected on a weekly basis, I linked the two datasets by assuming that the beliefs in a given week that lies between waves X and X+1 are equal to those of wave X, if the date of this week's Sunday is closer to the start date of wave X than to the start date of wave X+1, and equal to the beliefs of wave X+1 otherwise.

In all models, I am interested in the coefficient $\rho$, as it tells us whether the FPÖ vote share conditional on the corona populist turn of the FPÖ has an impact on the dependent variable, i.e. cases or deaths. It is important to note that this model does not allow an interpretation about whether increased support for the FPÖ predicts an increase in cases or deaths *in total*, as the time-invariant effect of the FPÖ-vote share is dummied out.[5]

*2.2.2 Aggregate-level results*

This subsection presents the results of the models on the number of i) cases per 1k inhabitants, and ii) deaths per 100k inhabitants.

Table 5 shows the results of the regressions predicting the number of cases per 1,000 inhabitants for different models. The standard errors are clustered by weeks and districts. The coefficient of interest, i.e. the effect of the interaction term between the policy switch dummy and the FPÖ vote share, is not statistically significant in any model. Naturally, the lagged dependent variable is positive and significant in any model, as Covid-19 cases produce new Covid-19 cases.

The first model includes the lagged cumulative number of cases per 1k inhabitants, as well as the lagged number of deaths as control variables. The lagged number of deaths is significant and negative, which suggests that people react to an increased perceived danger posed by Covid-19 by changing their behavior, i.e. by practicing social distancing. Unexpectedly, we do not find a significant "herd immunity" effect from the (lagged) number of cumulated previous cases.

Models (2) and (3) introduce time-varying district-specific belief indices as imputed by linking the district-data with survey data from the Austrian Corona Panel Project (Kittel et al. 2020, 2021) based on demographic characteristics as described above. Two indices are significant: the imputed belief that Covid-19 poses a high danger to public health reduces the number of cases, whereas the imputed belief that Covid-19 poses a high danger to the economy increases the number of cases. This is interesting, as the variation introduced by matching the time and district-level data with the ACPP data captures belief dynamics beyond the current state of the pandemic (which are captured by the time fixed effects) or demographic characteristics (which are captured in the district fixed effects).

[5] Table H3 and H4 in the online appendix are set up without district fixed effects and suggest that there is no significant effect of the FPÖ vote share prior to the policy switch.

Causality between health-related beliefs and infections and deaths is intricate: On the one hand, people who are more exposed to the virus may rationally believe that they are more at risk, hence increasing the number of infections. On the other hand, people who are more risk-aware may be more inclined to practice social distancing, wear masks, get vaccinated etc., hence reducing the number of infections. The regression results suggest that the infection-reducing effect prevails with regard to the number of cases.

The second finding, i.e. that the belief that Covid-19 poses a high danger to the economy increases the number of cases, is more straightforward: if the economic damage dealt by the Covid-19 crisis is primarily conceived as stemming from voluntary or government-mandated social distancing, individuals may consider social distancing as a trade-off. This was already modeled, e.g., by Allcott et al. 2020 in a utility maximization framework and by Proaño and Makarewicz in a behavioral model. Furthermore, I showed in the previous subsection that the belief that Covid-19 poses a high danger to the economy is positively associated with the level of skepticism, and the current results suggest that this relation also holds at the wider level.

Models (4)-(7) then also account for vaccinations, which have played a role in Austria since the beginning of 2021. Model (4) includes the cumulative number of second dose vaccinations (which is only available on the state-level as a time series). As expected, the respective coefficient is significant and negative due to an increase in herd immunity. Models (5)-(7) differ from model (4) by imputing a district-level vaccination timeline that is created based on cross-sectional data on the district-level distribution of vaccinations from the 21st of February 2022 and assuming that this share stays constant, thus matching it with the state-level time series readily available. While this imputation does not come without costs, as the relative distribution of vaccines may have changed over time, this imputation increases the significance and absolute value of the respective coefficient, hence providing credibility to the mechanism.

Models (6) and (7) then also control for the imputed belief indices. While there are no notable changes for model (7), a belief in high risks for the economy as a whole is not significant anymore in model (6). This result suggests that, as one might have expected, the impact of corona "skepticism" is at least partly transmitted through the channel of vaccination rates.

Table 5: Cases per 1k inhabitants

| | Without vaccinations | | | | With vaccinations | | |
|---|---|---|---|---|---|---|---|
| | **(1)** | **(2)** | **(3)** | **(4)** | **(5)** | **(6)** | **(7)** |
| lag(cases per 1k, 1) | 0.786*** | 0.784*** | 0.786*** | 0.782*** | 0.779*** | 0.777*** | 0.779*** |
| | (0.065) | (0.065) | (0.064) | (0.065) | (0.064) | (0.064) | (0.064) |
| lag(deaths per 100k, 1) | -0.013* | -0.012* | -0.013* | -0.012* | -0.012* | -0.011* | -0.012* |
| | (0.005) | (0.005) | (0.005) | (0.005) | (0.005) | (0.005) | (0.005) |
| FPÖ vote share x Policy switch | 0.008 | 0.005 | 0.003 | 0.006 | 0.005 | 0.002 | -0.001 |
| | (0.007) | (0.008) | (0.008) | (0.007) | (0.007) | (0.008) | (0.008) |
| Cumulative number of cases per 1k inhabitants | 0.000 | 0.000 | 0.000 | -0.001 | -0.004 | -0.005 | -0.004 |
| | (0.003) | (0.003) | (0.003) | (0.002) | (0.002) | (0.002) | (0.002) |
| lag(State-level second-dose vaccinations per 1k, 4) | | | | -0.003* | | | |
| | | | | (0.001) | | | |
| lag(Imputed district-level second-dose vaccinations per 1k, 4) | | | | | -0.004** | -0.004** | -0.004** |
| | | | | | (0.001) | (0.001) | (0.001) |
| Index: government measures are too lax | | -0.308 | | | | 0.748 | |
| | | (1.505) | | | | (1.125) | |
| Index: Covid-19 poses a high danger to personal health | | 0.654 | | | | -0.149 | |
| | | (1.800) | | | | (1.833) | |
| Index: Covid-19 poses a high danger to public health | | -15.145** | | | | -14.265** | |
| | | (4.624) | | | | (4.362) | |
| Index: Covid-19 poses a high danger to personal economic situation | | 1.272 | | | | 9.588 | |
| | | (6.192) | | | | (5.314) | |
| Index: Covid-19 poses a high danger to the economy | | 10.914** | | | | 6.404 | |
| | | (3.919) | | | | (3.608) | |
| Index: government measures are exaggerated | | | 3.274 | | | | 3.527 |
| | | | (3.938) | | | | (3.807) |

Table 5: Cases per 1k inhabitants

| | Without vaccinations | | | With vaccinations | | | |
|---|---|---|---|---|---|---|---|
| | **(1)** | **(2)** | **(3)** | **(4)** | **(5)** | **(6)** | **(7)** |
| Index: Covid-19 poses a low danger to personal health | | | 1.800 | | | | 3.369 |
| | | | (3.110) | | | | (3.115) |
| Index: Covid-19 poses a low danger to public health | | | 5.617 | | | | 4.170 |
| | | | (4.845) | | | | (4.709) |
| Index: Covid-19 poses a low danger to personal economic situation | | | -1.029 | | | | -6.491 |
| | | | (5.022) | | | | (4.994) |
| Index: Covid-19 poses a low danger to the economy | | | -7.336 | | | | -3.246 |
| | | | (7.826) | | | | (6.582) |
| Num.Obs. | 9212 | 9212 | 9212 | 9212 | 9212 | 9212 | 9212 |
| R2 | 0.929 | 0.929 | 0.929 | 0.929 | 0.929 | 0.930 | 0.929 |
| R2 Adj. | 0.927 | 0.927 | 0.927 | 0.927 | 0.928 | 0.928 | 0.928 |
| R2 Within | 0.602 | 0.603 | 0.603 | 0.604 | 0.607 | 0.608 | 0.607 |
| R2 Within Adj. | 0.602 | 0.603 | 0.602 | 0.604 | 0.606 | 0.607 | 0.607 |
| AIC | 18638.6 | 18621.5 | 18643.6 | 18599.6 | 18537.8 | 18519.2 | 18541.1 |
| BIC | 18674.2 | 18692.8 | 18714.9 | 18642.3 | 18580.6 | 18597.7 | 18619.5 |
| RMSE | 0.67 | 0.66 | 0.66 | 0.66 | 0.66 | 0.66 | 0.66 |
| Std.Errors | by: GKZ & time | by: GKZ & time | by: GKZ & time | by: GKZ & time | by: GKZ & time | by: GKZ & time | by: GKZ & time |
| FE: GKZ | X | X | X | X | X | X | X |
| FE: time | X | X | X | X | X | X | X |

* p < 0.05, ** p < 0.01, *** p < 0.001

Table 6 shows the results of the panel regressions predicting the number of deaths per 100,000 inhabitants. The coefficient of interest, i.e., the effect of the interaction term between the policy switch dummy and the FPÖ vote share, is positive and statistically significant at the 0.1% level for any model. This result suggests that the corona populist turn of the FPÖ did have an impact on the regional distribution of deaths in Austria in the sense that the number of Covid-19 deaths per capita after the policy switch are positively correlated with the vote share of the FPÖ at the last national elections before Covid-19 (i.e. 2019).

Every model also includes the lagged dependent variable (i.e. the number of deaths per 100k inhabitants in the previous week), which is significant and positive. While deaths do not necessarily produce new deaths (save for a Zombie apocalypse), the lagged dependent variable is a proxy for other important factors with regard to the evolution of the pandemic such as infections, ICU capacities etc.

Models (2) and (3) also feature the imputed belief indices as described above. Two indices are significant: First, the belief that Covid-19 poses a high danger to the personal economic situation is positive. This belief can be seen as a proxy for a certain type of skepticism (as reasoned above), and accordingly increases the number of deaths. Second, the belief that Covid-19 poses a high danger to personal health, which is negative. As reasoned above, the causality between health beliefs on the one side, and infections and deaths on the other side is intricate, as these beliefs may on the one hand represent rational perceptions, but on the other hand likely also trigger a behavioral response which may mitigate or even counteract any rational perception. The fact that this coefficient is negative seems to point to the fact that the latter effect prevails for this belief dimension.

The fact that the vote share of the FPÖ is strongly correlated with deaths after the policy switch, but not with cases per capita, seems at first glance to be paradoxical and to sow doubt on the hypothesis that the corona populist turn of the FPÖ contributed to the spread of the virus. However, Covid-19 tests in Austria have largely been conducted on individuals who self-reported their symptoms or who are named as being close contacts. Thus, they have in one way or another often been voluntary, in particular before the introduction of compulsory tests needed for certain activities in the beginning of February 2021, which means that there may have been a self-selection bias in testing.[6]

We can hypothesize that people who underestimate the virus (the "corona skeptics") or even deny the existence of the virus (the so-called "corona deniers") are less likely to report an infection and to name contacts. In this case, the number of *deaths per reported infection* in such communities would be higher.

In order to conduct a first test on this hypothesis, models (4)-(6) also control for the number of cases per capita in the two preceding weeks. Holding cases constant, the mortality rate may diverge for two reasons: first, due to a higher true infection fatality rate, given by, e.g., different age structures of the population (or, more precisely, the infected share of the population), and second due to a higher *perceived* infection fatality rate given by a higher share of undetected cases (i.e. a higher dark figure). It is interesting to see that the coefficient of interest only changes marginally. Since the age structure of the population did not change during the policy switch, this result hints to the fact that the policy switch indeed increased the share of undetected cases, i.e. the dark figure. This result would also explain the non-result regarding an impact on the number of cases: While the policy switch did not have an impact on the number of *reported* cases, there is reason to believe that it did have an impact on the *true* number of cases.

[6]From the beginning of February 2021 to the beginning of March 2022, either Covid-19 tests or a "green pass", i.e. proof of testing, vaccination or immunity due to a previous infection, were required for a varying number of activities. However, much of the testing regime relied on the use of self-tests that a) offer relatively low sensitivity and b) that can easily be manipulated. Furthermore, the largest relative increase in mortality with regard to the FPÖ vote share seems to have happened before February 2021, as can be seen in fig. 6.

I further explore the existence of such a testing bias empirically in subsection 2.3, and explore whether such a testing-bias is indeed able to explain the puzzling result that I do find a significant impact of the policy switch on the regional distribution of deaths, but not reported cases, in a stylized epidemiological model in section 3.

Once controlling for cases, the belief index regarding a high danger to public health becomes significant. The fact that this coefficient is positive, while the coefficient regarding a high danger to personal health remains significant and negative, seems to showcase the intricate causality regarding health beliefs and mortality as reasoned above.

Table 6: Deaths per 100k inhabitants

| | **Without controlling for cases** | | | **Controlling for cases** | | |
|---|---|---|---|---|---|---|
| | **(1)** | **(2)** | **(3)** | **(4)** | **(5)** | **(6)** |
| lag(deaths per 100k, 1) | 0.401*** | 0.398*** | 0.400*** | 0.327*** | 0.323*** | 0.326*** |
| | (0.037) | (0.037) | (0.037) | (0.037) | (0.037) | (0.037) |
| FPÖ vote share x Policy switch | 0.104*** | 0.089** | 0.096*** | 0.096*** | 0.082*** | 0.095*** |
| | (0.020) | (0.028) | (0.022) | (0.017) | (0.025) | (0.019) |
| Cases per 1k inhabitants in the weeks t-1 and t-2 | | | | 0.321*** | 0.327*** | 0.322*** |
| | | | | (0.056) | (0.055) | (0.056) |
| Index: government measures are too lax | | 14.219 | | | 22.442 | |
| | | (14.552) | | | (13.628) | |
| Index: Covid-19 poses a high danger to personal health | | -28.494* | | | -34.096** | |
| | | (11.809) | | | (12.479) | |

Table 6: Deaths per 100k inhabitants

| | Without controlling for cases | | | Controlling for cases | | |
|---|---|---|---|---|---|---|
| | **(1)** | **(2)** | **(3)** | **(4)** | **(5)** | **(6)** |
| Index: Covid-19 poses a high danger to public health | | 21.553 | | | 46.920** | |
| | | (17.276) | | | (17.641) | |
| Index: Covid-19 poses a high danger to personal economic situation | | 50.348** | | | 37.716* | |
| | | (17.404) | | | (17.287) | |
| Index: Covid-19 poses a high danger to the economy | | 16.699 | | | -1.214 | |
| | | (13.902) | | | (12.915) | |
| Index: government measures are exaggerated | | | -13.617 | | | -3.820 |
| | | | (10.558) | | | (10.735) |
| Index: Covid-19 poses a low danger to personal health | | | 19.383 | | | 11.476 |
| | | | (10.153) | | | (10.318) |
| Index: Covid-19 poses a low danger to public health | | | -8.387 | | | -26.412 |
| | | | (16.312) | | | (15.405) |
| Index: Covid-19 poses a low danger to personal economic situation | | | -14.455 | | | 2.792 |
| | | | (18.783) | | | (19.304) |
| Index: Covid-19 poses a low danger to the economy | | | -40.515 | | | -32.912 |
| | | | (25.868) | | | (23.730) |
| Num.Obs. | 9212 | 9212 | 9212 | 9212 | 9212 | 9212 |
| R2 | 0.575 | 0.576 | 0.575 | 0.599 | 0.600 | 0.599 |
| R2 Adj. | 0.566 | 0.567 | 0.566 | 0.590 | 0.591 | 0.590 |
| R2 Within | 0.157 | 0.159 | 0.158 | 0.204 | 0.207 | 0.205 |
| R2 Within Adj. | 0.157 | 0.158 | 0.157 | 0.204 | 0.206 | 0.204 |
| AIC | 43431.9 | 43419.9 | 43434.7 | 42905.5 | 42884.4 | 42908.1 |
| BIC | 43453.3 | 43476.9 | 43491.8 | 42934.0 | 42948.5 | 42972.2 |
| RMSE | 2.56 | 2.55 | 2.55 | 2.48 | 2.48 | 2.48 |
| Std.Errors | by: GKZ & time | by: GKZ & time | by: GKZ & time | by: GKZ & time | by: GKZ & time | by: GKZ & time |
| FE: GKZ | X | X | X | X | X | X |
| FE: time | X | X | X | X | X | X |

* p < 0.05, ** p < 0.01, *** p < 0.001

### *2.2.3 Aggregate-level robustness checks*

Appendices D-I show the results of several robustness checks. The first robustness check concerns a potential endogeneity due to heterogeneous evolution of the pandemic prior to the policy switch. The overview of cases and deaths shown in figure 6 suggests that those 50% of the districts, in which the FPÖ achieved its strongest outcomes recorded less deaths during the first, but more deaths in particular during the second wave. Thus, one could hypothesize that the underlying mechanism driving the results is in fact not a political, but an epidemiological one: if more people became infected and/or died during the first wave, the subsequent waves could be milder either due to herd immunity or due to increased awareness of the danger of the virus, and the FPÖ vote share could merely correlate with this underlying mechanism. I explore this hypothesis in appendix D and show that my results are robust to including an interaction term of the intervention with either the number of deaths or the number of cases before the intervention as an additional control variable, even though such a counteractive effect indeed exists and the magnitude of the coefficient of interest is reduced.

In a related concern, the social share of the population affected by the coronavirus may differ between the first and the subsequent waves, e.g., the second wave may have hit more elders than the first wave, thus increasing the case fatality rate. I thus include interaction terms between various demographic groups and the policy switch to explicitly account for this potential demographic endogeneity instead of relying on the belief indices. Appendix E shows that my results are robust to such an inclusion, although the population above the age of 64 seems to have been hit harder in the infection waves following the first one (i.e. after the policy switch).

In the next set of robustness checks, presented in appendix F, I check whether the results change when I include interaction terms between the policy switch and the vote share of other parties. All results are robust to the inclusion of interaction terms with any other party. Only a few additional results are significant for the number of deaths per 100k inhabitants, namely the interaction term between the policy switch and the vote share of a) the social democratic party (SPÖ), where it is positive, b) the major governmental party (ÖVP), where it is negative, as well as c) the combined vote share of minor parties, where it is positive in some models. It is important to note, however, that these results are not robust to the robustness checks introduced in appendix D and thus are likely an artifact of epidemiological considerations.

Alashoor et al. (2020) study social distancing behavior in the United States using a survey and show that, for their respondents, not partisanship, but their vote in the presidential elections 2016 mattered for compliance with social distancing measures. Respondents who voted for Trump in 2016 were less likely to follow social distancing rules, if their attitude towards social distancing was negative. The last Austrian presidential elections before the Covid-19 crisis were also held in 2016, where the candidate of the FPÖ, Norbert Hofer, lost in the run-off against the candidate of the Greens, Alexander van der Bellen. In the next robustness check, presented in appendix G, I hence test whether including an interaction term between the vote share for Norbert Hofer and the policy switch influenced the results. Interestingly, this exercise even increases the effect size for the interaction term of the FPÖ and the policy switch with regard to the number of deaths, whereas the coefficient for the new interaction term is not significant.

In the next robustness check, presented in appendix H, I use alternative model specifications in order to explore the robustness of my results in the face of the dynamic panel bias (Nickell 1981), which comes into play for fixed effects models that include an autoregressive term where the underlying data has a small t and large n. Since t>n in my panel, the bias should not be of a great concern. Nevertheless, I estimate i) static panel models (tables E1 and E2), i.e. models without an autoregressive term, as well as ii) models without district fixed effects, but include the autoregressive term. This procedure is suggested by the literature (see, e.g., Angrist and Pischke 2009) because it helps to establish boundaries of the true effect, since static panel models will overestimate the coefficient, whereas pooled models will underestimate it. My results are found in all of those models.

In the final robustness check, presented in appendix I, I compute the models estimating cases and deaths per capita including the regional time-specific belief indices imputed with the ACPP data, but without the FPÖ x policy switch interaction term. In contrast to the baseline specification, the belief index "Covid-19 poses a low danger to personal health" significantly increases the number of deaths per capita. This is important, as I showed in section 2.1 that this belief significantly increases among people who voted for the FPÖ in 2019 after the switch.

### *2.3 Individual-level evidence from the Covid-19 prevalence study in November 2020*

My analysis of the aggregate-level data provides a puzzling result, as it suggests that the FPÖ policy switch has affected the regional distribution of deaths, but not of cases per capita. I hypothesized that this is due to a self-selection bias in testing that causes a correlation between the FPÖ vote share and the share of undetected cases, and found some evidence for this hypothesis in the aggregate-level data. This subsection now aims to establish a link between "Corona skepticism" and the true infection status, as well as the dark figure, on the individual level.

While there is no Austrian individual-level data source that combines partisan affiliation with infection status, the Covid-19 prevalence study conducted in November 2020 (Paškvan et al. 2021) provides information about the individuals' stance on Covid-19 containment policy, true infection status (determined with a PCR test), past infection status (determined with an antibody neutralization test) and reported infection status. This study was conducted by the Austrian statistical office (Statistik Austria) and its participants were randomly chosen to form a statistically representative sample.

These data thus allow to test the hypotheses that "corona skeptics", i.e. people who are opposed to containment policies, are i) more likely to contract Covid-19 and ii) less likely to test themselves.

Similar to the Austrian Corona Panel Project, the prevalence study includes a Likert scale survey item regarding the individual's stance to the containment policies ranging from 1 ("definitely exaggerated") to 5 ("definitely insufficient"). In order to facilitate the statistical analysis, I transform this scale to a dummy variable "corona skeptical", where a 1 covers the views that the policy measures against Covid-19 are "definitely exaggerated" or "rather exaggerated", whereas 0 implies that the individual thinks that the measures are "suitable", "rather insufficient" or "definitely insufficient".[7]

I then investigate the relationship between corona skepticism and four variables: a) reported infection (i.e. was officially known to be Covid-19 positive at or prior to the time of the survey), b) PCR test positive (i.e. tested positive for Covid-19 during this survey), c) antibody test positive (i.e. a blood test revealed a neutralizing level of antibodies), d) unreported current infections (i.e. the PCR test was positive during this survey, but the infection was not officially known at that time). I only analyze individuals for which at least a PCR test result (n=2290) and a policy stance (2283 out of the 2290) are available.

[7] There is a crucial difference between the prevalence study of November 2020, which was conducted during an infection peak, and the Austrian Corona Panel Project data, which also covers periods during which the danger posed by Covid-19 has been objectively low (e.g. summer of 2020 or summer of 2021). Hence, I do not use the same terminology for the dummy variable used in section 2.1 and this section, even though the survey items on which the two dummy variables are based, are phrased highly similarly.

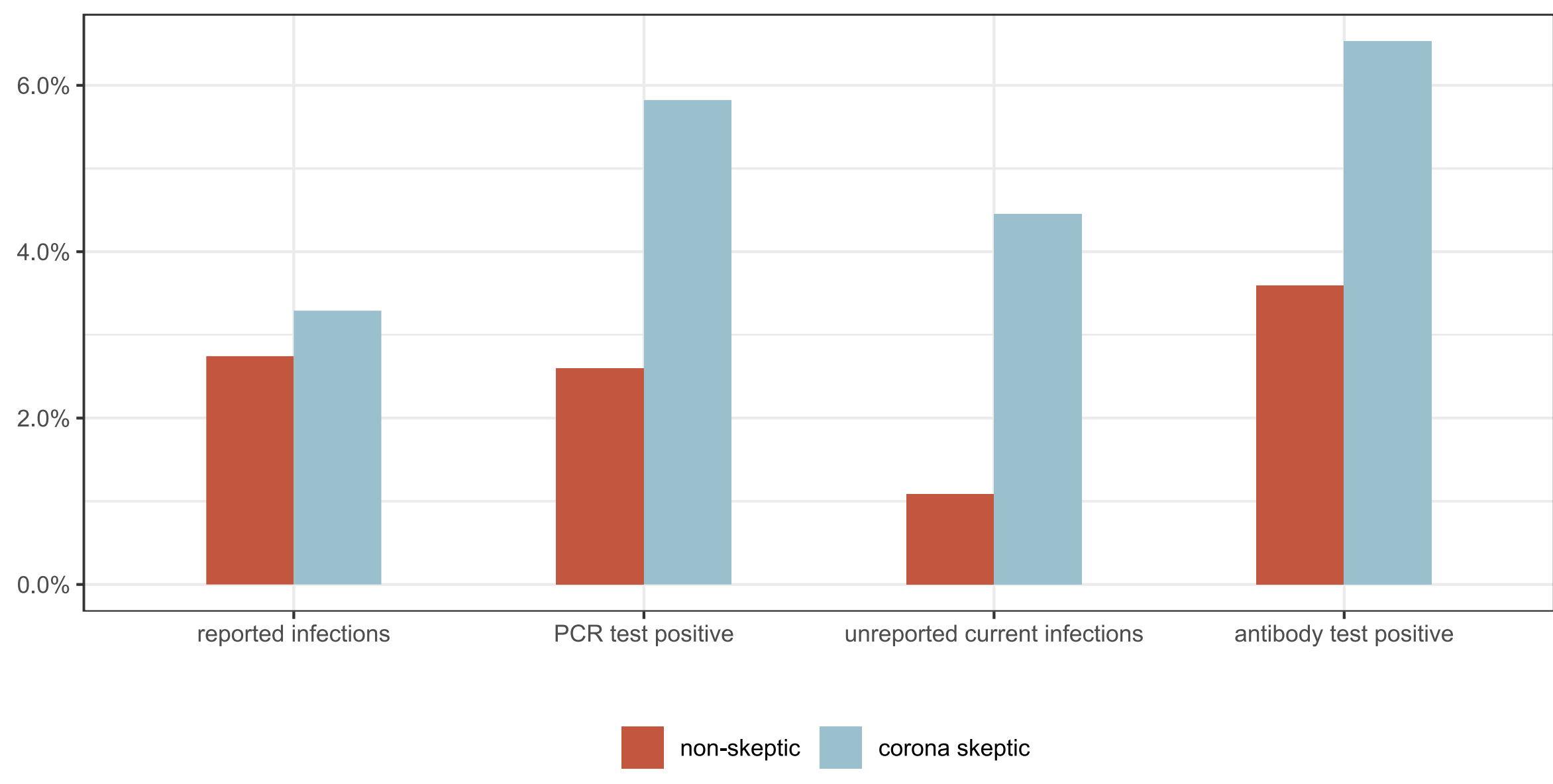

**Figure 8**: Epidemiological state for corona skeptics and non-skeptics in November 2020

Figure 8 suggests that there is no large difference between corona skeptics and non-skeptics with regard to the reported infections, but a large gap with regard to unreported infections. These results are fully in line with the self-selection bias hypothesis.

In order to test whether the graphical intuition holds, I estimate models using probit regressions and the following specification:

$$y_i = s_i\beta + \alpha + \varepsilon_i \tag{6}$$

Where $y_i$ is a binary variable describing whether a) a positive PCR/antibody test has been conducted or b) individuals were already known to be infected prior to the study, $s_i$ is a binary variable describing whether an individual is "corona skeptical" or not as discussed above, $\beta$ is the coefficient of interest, $\alpha$ is the intercept and $\varepsilon_i$ is an error term.

Table 7 shows the results. The first model suggests that corona skeptical individuals have an increased chance to test positive at the 1% significance level within the whole population (i.e. also including people who are officially known to be infected). The second model shows that corona skeptical individuals are even more likely to test positive than non-skeptical individuals within the subpopulation that only includes people who are not officially known to be infected. The third model shows that corona skepticism does not predict a significant increase in the official infection status. Jointly, these results show that corona skepticism is indeed positively correlated with the dark figure on the individual level. Finally, the fourth model shows that corona skeptical individuals were more likely to have gone through a past infection. Again, all of these results support the self-selection bias hypothesis. Furthermore, the fact that there is no significant difference between regarding reported infections seems to mirror the aggregate-level results regarding the relationship between the number of reported infections per capita and the FPÖ vote share.

Table 7: Individual-level evidence from the Covid prevalence study in November (Probit)

| | **DV: PCR test positive** | | **DV: Reported infection** | **DV: Antibody test positive** |
|---|---|---|---|---|
| | **(1) Group: All individuals** | **(2) Group: Only unreported** | **(3) Group: All individuals** | **(4) Group: All individuals** |
| Corona skeptical | 0.374** | 0.595*** | 0.079 | 0.288* |
| | (0.118) | (0.140) | (0.136) | (0.113) |
| Intercept | -1.944*** | -2.296*** | -1.918*** | -1.800*** |
| | (0.061) | (0.084) | (0.059) | (0.055) |
| Num.Obs. | 2283 | 2225 | 2283 | 2219 |
| AIC | 634.0 | 363.8 | 594.5 | 757.4 |
| BIC | 645.5 | 375.2 | 605.9 | 768.8 |
| Log.Lik. | -314.995 | -179.883 | -295.228 | -376.689 |
| F | 10.041 | 17.940 | 0.334 | 6.438 |
| RMSE | 0.53 | 0.40 | 0.51 | 0.58 |

* $p < 0.05$, ** $p < 0.01$, *** $p < 0.001$

## 3 Theory: Insights from a heterogeneous mixing SIR model

This section is devoted to understanding whether the proposed solution to the puzzling aggregate-level result from the empirical section, namely that the FPÖ policy switch increased deaths, but not necessarily reported cases due to biased testing, holds in a simple theoretical epidemiological framework. In order to do so, I extend the classical SIRD model (Kermack and McKendrick 1927) in a twofold way:

1.) I add a quarantined compartment denoted by Q that only includes detected active cases. A certain fraction of infected is assumed to test themself upon infection and is then quarantined, i.e. their social contacts are set to 0. I further assume that all critical cases are detected, since they seek medical attention and get tested for showing symptoms of Covid-19. Accordingly, only people in the quarantine compartment may die. Holding constant the fraction of infected who will eventually die (i.e. the true infection fatality rate), the fraction of quarantined who die (i.e. the reported infection fatality rate) depends on the fraction of non-critical cases who opt to get tested voluntarily, i.e. on the fraction of critical cases in the quarantine compartment.

2.) I split the compartments governing the susceptible, the infected and the quarantined to incorporate two different groups: one group showing low compliance (the corona skeptics) and another showing high compliance (the majority). I consider differences in a) social distancing, and b) the propensity to get tested. I also consider the case of homophilic mixing, i.e. that individuals of a certain group are more likely to get into contact with members of their own group than members of the other group (which is why I need two different compartments for the infected).

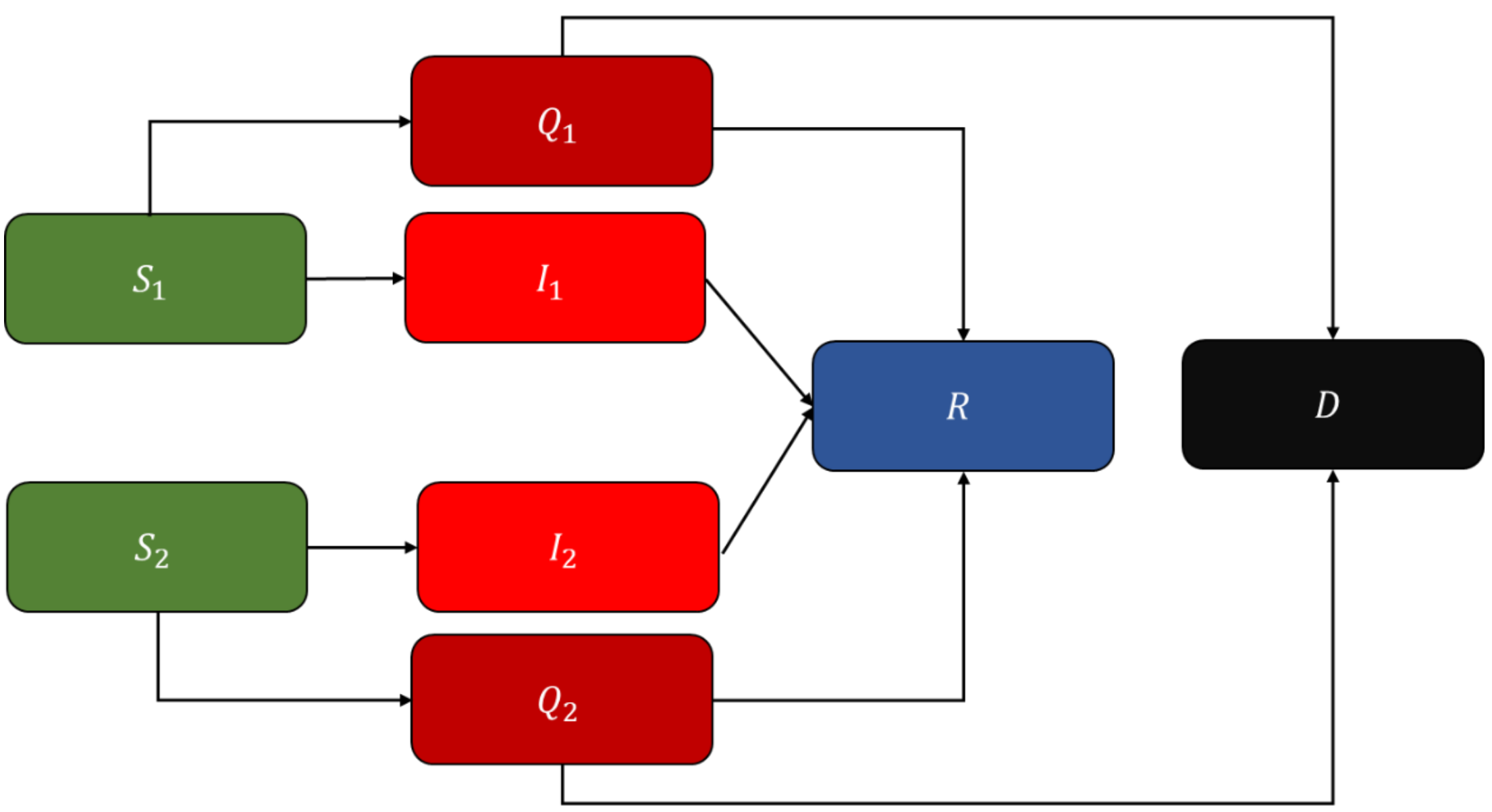

**Figure 9**: Depiction of the compartments

In setting up the laws of motion between the different compartments (see fig. 9), I largely follow the preferred mixing model described by Brauer (2008), which in turn largely follows

Nold (1980).[8] In contrast to comparable models such as the homophilic mixing model proposed by Ellison (2020), the model used in my paper is able to replicate the standard homogenous mixing model as a special case if the behavior of the two types of agents (especially the basic reproduction numbers $R_{01}$ and $R_{02}$ respectively) is equal. The model is most closely related to the one proposed by Bai and Brauer (2021), who also study a SEIR model populated by two types of agents who differ in their basic reproduction number and add a quarantined compartment. In contrast to their model, however, my model also considers differences in testing between the two groups (to account for the Austrian empirics), as well as the case of homophilic mixing. Their model, on the other hand, also features an exposed compartment in order to capture pre-symptomatic infections.

The laws of motion between the different compartments in my model are given as follows, where $S_i$ denotes the susceptibles of group i, $I_i$ the infectious, $Q_i$ the quarantined, $N_i$ the size of group i at period 0, $\beta_i$ the number of infectious contacts from a member of group i, $h$ the homophily of social contacts, $\alpha_i$ the propensity to get tested, $R_{0i}$ the basic reproduction number of group i, $\frac{1}{\gamma}$ the duration the illness, and $\mu_i$ the fraction of detected cases who eventually die:

$$\dot{S}_i(t) = -S_i(t)\beta_i\left(p_{i1}\frac{I_1(t)}{N_1} + p_{i2}\frac{I_2(t)}{N_2}\right) \tag{7}$$

$$\dot{I}_i(t) = (1-\alpha_i)S_i(t)\beta_i\left(p_{i1}\frac{I_1(t)}{N_1} + p_{i2}\frac{I_2(t)}{N_2}\right) - \gamma I_i(t) \tag{8}$$

$$\dot{Q}_i(t) = \alpha_i S_i(t)\beta_i\left(p_{i1}\frac{I_1(t)}{N_1} + p_{i2}\frac{I_2(t)}{N_2}\right) - \gamma Q_i(t) \tag{9}$$

$$\dot{R}(t) = \gamma\big(I_1(t) + I_2(t)\big) + \gamma\big((1-\mu_1)Q_1(t) + (1-\mu_2)Q_2(t)\big) \tag{10}$$

$$\dot{D}(t) = \gamma\big(\mu_1 Q_1(t) + \mu_2 Q_2(t)\big) \tag{11}$$

[8] Brauer (2008) considers the fraction that each group *currently* makes up as part of the total population. I refrained from implementing this logic in order to retain the classical SIR outcome as a special case.

Where

$$p_{ij} = \begin{cases} h + (1-h)\, p_j & if\ j = i \\ (1-h)\, p_j & if\ j \neq i \end{cases} \tag{12}$$

$$p_i = \frac{(1-h)\gamma R_{0i} N_i}{(1-h)\gamma R_{01} N_1 + (1-h)\gamma R_{02} N_2} \tag{13}$$

$$\beta_i = \gamma R_{0i} \tag{14}$$

In order to better disentangle the effects of behavioral differences of the two groups, I make the following practical assumption: Individuals of both groups are equally likely to die as a result of an infection with a probability of $\pi$. We can thus set the probability that a quarantined person dies at $\mu_i = \frac{\pi}{\alpha_i}$ and set a lower boundary for $\alpha_i$, as I assumed previously that at least all critical cases are tested, i.e. $\alpha_i \geq \pi$.

### *3.1 Homogenous mixing*

Let us first consider the case of homogenous mixing, i.e. $h = 0$ and $R_0 = R_{01} = R_{02}$. In this case, we can immediately see that $p_{ij} = p_i = \frac{N_i}{N_1+N_2}$. Thus, if we normalize the population to 1, i.e. $N_1 + N_2 = 1$, the dynamic governing the susceptibles collapses to the dynamic of the classical one-group SIR framework, i.e.:

$$\dot{S}_i(t) = -S_i(t)\gamma R_0 \big(I_1(t) + I_2(t)\big) \tag{15}$$

In such a case, differentiating between two infectious compartments is unnecessary (see proposition 1).

**Proposition 1**: Suppose two groups who mix homogeneously and only differ with regard to their propensity to get tested. Such a difference can only affect both groups equally (relative to their share) in terms of deaths or the sum of infected and quarantined.

**Proof**: Differentiating eq. (15) with regard to I1 and I2 yields the same results, hence the relative share of $I_i$ as part of the total infected does not have an impact on the evolution of $S_i$:

$$\frac{\partial \dot{S}_i(t)}{\partial I_1} = \frac{\partial \dot{S}_i(t)}{\partial I_2} = -S_i(t)\gamma R_0 \tag{16}$$

Nevertheless, we could use a homogenous mixing framework to consider differences in the propensity to get tested (and subsequently get quarantined), i.e. a different evolution of $Q_1$ and $Q_2$.

As is common in this stream of literature (e.g. Acemoglu et al. 2021), I support my arguments using numerical simulations and sensitivity analyses. The basic reproduction number of the wild type of the SARS-CoV-2 virus causing Covid-19 (i.e. before any viral mutations that may alter viral fitness, see Mellacher (2022)) is typically estimated to range from 2.5-3.5 (e.g. Zhao et al. 2020).

To capture political reasons for differences in social distancing, I concentrate on a time when social distancing is at least encouraged. Thus, I will concentrate on values of $R_0$ between 3 and 1 (where the number of infected could not grow exponentially). While the choice of $\gamma$ plays a crucial role in predicting the spread of the virus and determining an optimal containment policy (Bar-On et al. 2021), it is not important in this stylized model that only seeks to explain certain features of the pandemic. For simplicity, I assume that $\gamma = 1/7$, implying that infected recover or die on average in one week. I further set $\alpha_1 = 0.45$ and $\alpha_2 = 0.27$, in line with the empirical data from Austria as presented in subsection 2.3. I report the results of simulations for differing basic reproduction numbers and shares of skeptics in fig. 10, and synthesize them in result 1.

**Result 1**: Suppose two groups 1 and 2 who mix homogeneously and only differ with regard to their propensity to get tested, and subsequently isolated ($\alpha_1 > \alpha_2 > 0$). Then an increase in the share of group 2 $N_2$ has a monotonically increasing effect on the cumulative number of deaths (see fig. 6, right), while it may simultaneously b) decrease the cumulative number of reported infections. More precisely, I find a monotonic decrease in the number of reported infections if $R_0$ is high enough, and a monotonic increase if $R_0$ is low enough. There is a non-linear relationship between the size of $N_2$ and the reported infections for intermediate levels of $R_0$.

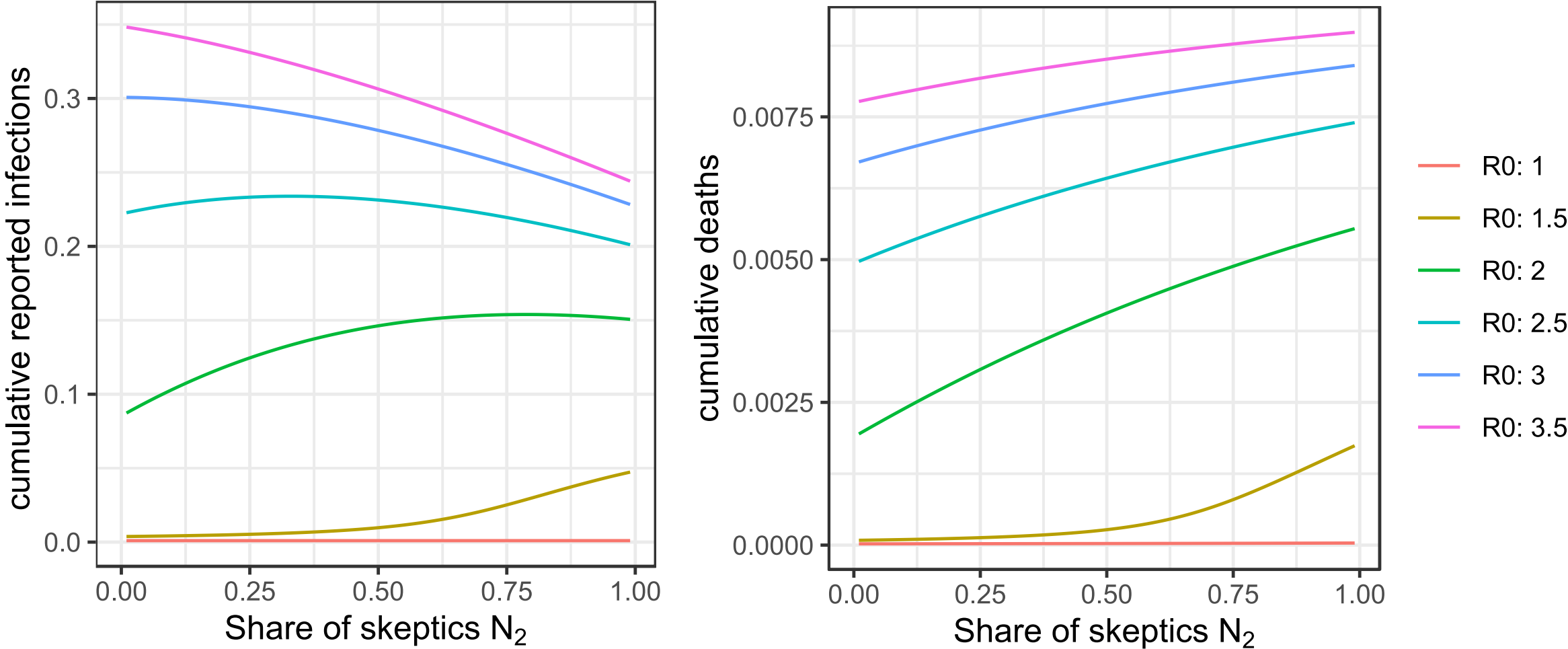

**Figure 10**: Cumulative number of reported infections (left) and deaths (right) after 500 periods. Each line in this figure shows the results of 100 distinct numerical simulations.

While result 1 shows that a self-selection bias in testing can indeed be sufficient to explain the aggregate-level phenomenon even in a simple homogenous mixing framework, this model has two apparent problems: First, proposition 1 shows that the total share of infected people cannot differ between the two groups *within* a given population (e.g. within a district), i.e. skeptics could not be affected in a different way than non-skeptics in a given district, which seems to be questionable. Second, this model assumes that there are no behavioral differences between the two groups except for the different testing behaviors. In order to address these issues, I extend my model to heterogeneous mixing in the following two subsections.

### *3.2 Proportionate mixing*

As soon as the two subpopulations engage in different activity patterns, i.e. $R_{01} \neq R_{02}$, homogenous mixing is implausible. If, for instance, group 1 only has one infectious contact per day, whereas group 2 has five, members of group 2 cannot on average have 2.5 infectious contacts with members of group 1, if the two groups are equal-sized. The specification by Brauer (2008), which provides the basis of my model, accounts for this fact. If activity patterns differ, but mixing is not homophilic, it is proportionate, i.e. members of a specific group meet members of another specific group according to their relative population shares and basic reproduction numbers as specified above. As a result, outcomes for both groups cannot be used interchangeably anymore. Instead, we must trace $I_1$ and $I_2$ separately.

**Result 2**: Suppose two groups 1 and 2 who mix proportionately and only differ with regard to their basic reproduction number ($R_{02} > R_{01} > 0$). Then an increase in a) the basic reproduction number of 2 $R_{02}$ or b) the share of group 2 $N_2$ increases the cumulative number of infected for both groups, but the increase (relative to their group size) is stronger for group 2.

The underlying simulations for result 2 are depicted in figure 11, which shows the share of susceptibles left in each subpopulation after 500 simulation periods (approximating the steady-state equilibrium). The top part shows the impact of a varying share of corona skeptics ($N_2$) in the population for 4 different levels of $R_{02}$. The simulations here are initialized with $R_{01} = 2.5$, (with the other parameters again set at $\gamma = 1/7$, $\pi = 0.01$, $\alpha_1 = 0.45$ and $\alpha_2 = 0.27$, as in the homogenous mixing model).

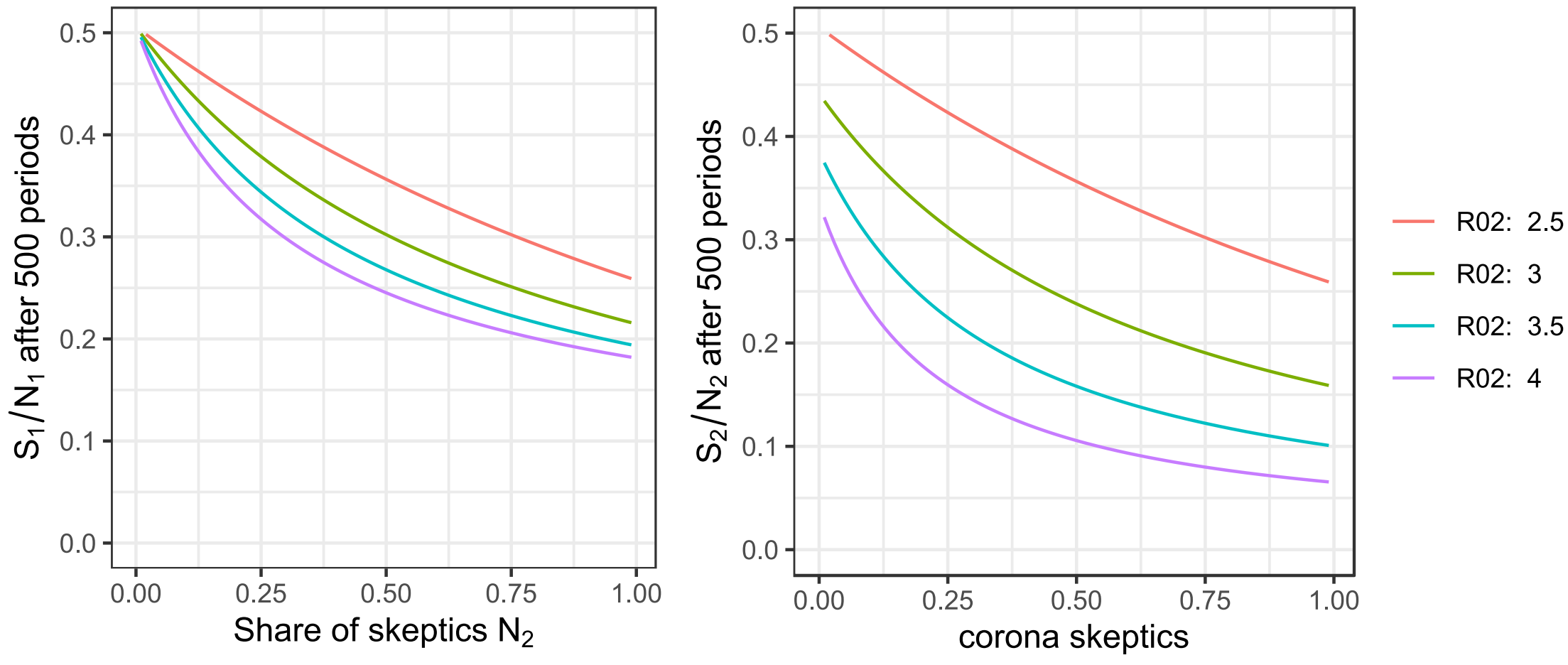

**Figure 11**: Susceptibles left of the non-skeptics (left) and the skeptics (right) divided by their initial share for $R_{01} = 2.5$ and varying $R_{02}$ and a varying share of corona skeptics under proportionate mixing

*3.3 Homophilic mixing*

Finally, I consider the case of homophilic mixing ($h > 0$), i.e., when “corona skeptics” are even more likely to meet each other than given by their relative basic reproduction numbers. Such a homophily in mixing could be created, for instance, by the anti-lockdown protests, which were frequently held in Austria. These settings are of particular interest, as they surely also increase the number of social contacts of the group of skeptics. Figure 12 shows simulation results for two different types of basic reproduction numbers for skeptics and a varying degree of homophily. Result 3 synthesizes the findings.

**Result 3**: Suppose two groups 1 and 2 who differ with regard to a) their respective basic reproduction number ($R_{02} > R_{01} > 0$) and b) their propensity to test themselves ($\alpha_1 > \alpha_2 > 0$) and who engage in homophilic mixing ($h > 0$). Let $f(h, N_2)$ be a function of the cumulative number of reported cases and $g(h, N_2)$ be a function of the cumulative number of deaths in the steady state. Then an increase in $h$ reduces the absolute value of $\frac{\delta^2 f}{\delta {N_2}^2}$ and $\frac{\delta^2 g}{\delta {N_2}^2}$ up to a point where it $\frac{\delta^2 f}{\delta {N_2}^2} = \frac{\delta^2 g}{\delta {N_2}^2} = 0$ for $h = 1$, i.e. $\frac{\delta f}{\delta N_2}$ and $\frac{\delta g}{\delta N_2}$ are constant. While $\frac{\delta g}{\delta N_2} > 0$, $\frac{\delta f}{\delta N_2}$ may be positive, negative or zero.

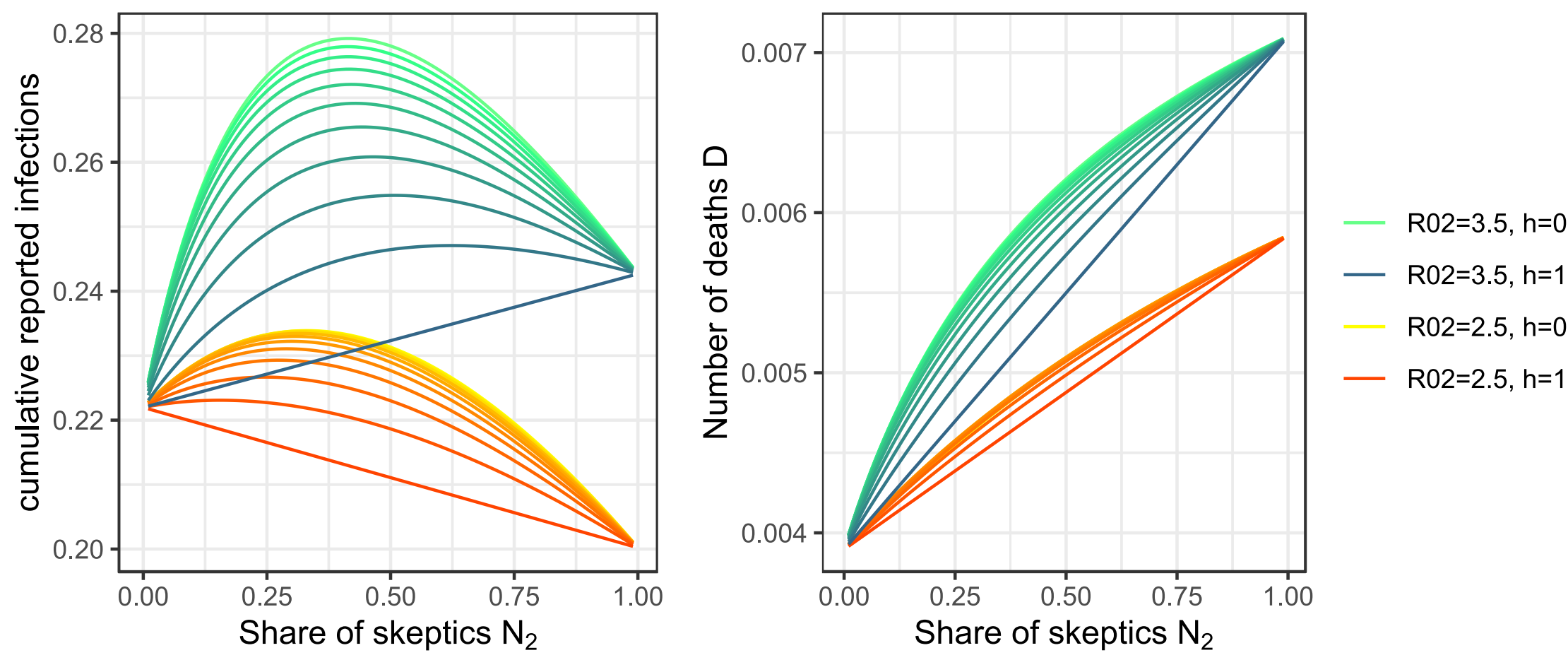

**Figure 12**: Cumulative reported infections (left) and deaths (right) after 500 periods for $R_{01} = 2.5$ and varying homophily $h$ as well as a varying number of corona skeptics ($N_2$).

It is obvious from result 3 that a situation in which the reported number of cases do not increase with $N_2$ can only be true in special cases, namely if $h = 1$ and $\frac{\delta f}{\delta N_2} = 0$. However, the simulation results suggest that a multitude of parameter settings exist for which we will not find a (linear) statistically significant relationship, in particular if we only observe a fraction of the possible values of $N_2$, as it seems to be true for the Austrian case.

## 4 Conclusion

In the beginning of this paper, I introduced two concepts: first, “corona skepticism”, which is a political view on the individual level which is “skeptical” of the danger posed by the Covid-19 virus and the need for containment measures. Second, and distinct from “skepticism”, I defined corona populism as a political strategy that is used by parties and candidates that aims to exploit and foster this “skepticism”. This definition is inspired by the definition of populism given by Acemoglu et al. (2013), who describe populism as a strategically chosen political platform that receives significant public support, although its adoption would cause adverse effects on the majority of the population.

I then theorized about the relationship between “corona skepticism” and “corona populism”, which is a special case of the relationship between party platform and voter preferences more generally. While the classical work by Downs (1957) suggested that parties set their platform in order to maximize their vote share (i.e. the party platform is determined by the voter preferences and not vice versa), more recent work (e.g. Bechtel et al. 2015; Druckman et al. 2013; Grewenig et al. 2020; Slothuus 2010) showed that voters tend to adapt their policy preferences to the platform of their party. I argued that the two explanations are not mutually excludable, but rather complementary and can hence be unified in a general theory of political opinion dynamics.

I then proceeded to test this theory at the example of a natural experiment given by a policy U-turn made by the Austrian major right-wing party FPÖ, which switched from a pro-lockdown to an anti-lockdown platform at the end of the first wave of Covid-19 infections.

First, I studied whether voters reacted to the turn by changing their voting intentions (“Downsian” effect) using data from the Austrian Corona Panel Project (Kittel et al. 2020, 2021). In order to minimize concerns about endogeneity that could be given by the joint presence of the “Downsian” effect and the “party identification” effect, I studied this question separately for survey respondents who claimed that they have voted for the FPÖ at the last national elections, which were held in 2019, i.e. before the Coronavirus crisis, and others. While the former group could have driven the platform of the party by themselves, the latter group is not suspected to exhibit a strong “party identification” effect.

My analysis showed that the support for the FPÖ among their voters in 2019 decreased following the turn. This decrease was mostly driven by those who believed that the government measures against Covid-19 were appropriate or too lax, while the support among those who were corona skeptical remained fairly constant. In contrast, the support for the FPÖ among "corona skeptics" who did not declare that they voted for the FPÖ in 2019 increased drastically. Jointly, these results indicate that the FPÖ policy turn i) was not mainly driven by a desire to satisfy their own voter base (or at least was not successful in doing so), and ii) induced voters to change their party preferences in order to align them with their policy views on Covid-19 (i.e. a "Downsian" effect exists).

Second, I studied whether FPÖ partisans realigned their policy preferences in order to match the platform propagated by their party, and whether any behavioral differences may have emerged due to this realignment (i.e. the "party identification" effect).

I investigated this question by studying i) individual-level beliefs and perceptions of FPÖ voters vs. others using a difference-in-differences approach with data from the Austrian Corona Panel Project, and ii) aggregate-level outcomes with regard to weekly reported Covid-19 infections and deaths on the district-level using two-way fixed effects panel models.

Before the policy switch, respondents who declared that they voted for the FPÖ in 2019 did not exhibit statistically significant differences with regard to their beliefs about personal or public health impact of the Covid-19 crisis. They were, however, significantly more likely to believe that it poses a high danger to the economy. FPÖ voters were furthermore more likely opposed against government policy even before the policy switch, although their opposition was stronger in both dimensions, i.e. they were more likely to believe that government policy was exaggerated and more likely to believe that it was too lax.

After the policy switch, however, respondents who voted for the FPÖ at the last national elections were less likely to believe that Covid-19 poses a high danger to public health, and more likely to believe that it poses a low danger to public and private health. In addition to that, those respondents coordinated their views on the appropriateness of government policy towards Covid-19, as the support for the view that government policy was exaggerated increased at the expense of the opposite view.

These findings support the “party identification” hypothesis which suggests that partisans tend to adjust their policy preference towards those of “their” political party. While the finding that partisan policy has an effect on voter beliefs is in line with previous findings on other policy views, my current findings are important as the Covid-19 crisis was a highly salient question that ultimately have amounted to a life vs. death situation for some of these voters.

Using the individual-level data set, I further showed that the belief that Covid-19 poses a high danger to the economy is linked to the belief that government policy against Covid-19 is exaggerated. I argued that this result stems from the fact that containment policy is often conceived as a trade-off between economic and health outcomes. Hence, people who are particularly concerned about the economic fallout of the crisis may tend to believe that the government puts too much emphasis on saving lives in their design of containment policy.

I then investigated whether the FPÖ policy switch could have had an effect on the regional distribution of infections and deaths in Austria based on the district-level support for the FPÖ at the last national elections using two-way fixed effects models with different specifications and a wide array of robustness checks. This exercise may tell us whether behavioral differences emerged from the realignment that was shaped by the policy turn.

In a subset of models, I combined the aggregate-level dataset with the individual-level data from the Austrian Corona Panel Project to impute regional beliefs about the Covid-19 pandemic and to account for a potential endogeneity with regard to “skeptical” beliefs. I showed that both imputed health perceptions and beliefs about the economic fallout are significantly correlated with Covid-19 cases and deaths per capita. While the causality between health-related and Covid-19 infections and deaths is intricate, because these effects likely capture (more or less rational) perceptions *and* counteractive behavioral responses, we can see that beliefs in high economic risks are associated with an increase in the number of cases, deaths and in the case fatality rate, suggesting that this belief is indeed a proxy for at least some kind of “corona skepticism”, as seen on the individual-level.

Regarding the impact of the corona populist turn, my regression analysis suggests that the FPÖ policy switch had a significant impact on the regional distribution of Covid-19 related deaths, but not of the regional distribution of (reported) infections based on an interaction term between the policy turn and the FPÖ vote share in the respective district. This result is puzzling, as nobody can die from Covid-19 without contracting it first. I hypothesized that the solution to this puzzle can be found in a self-selection bias inherent to the Austrian containment policies, in particular before the introduction of mandatory Covid-19 testing in February 2021: The policy stance of the FPÖ caused their voter base to take the virus less seriously, who then did not only practice less social distancing, but also reported their symptoms less often, which means that they were less likely tested. Hence, the true infection rate among FPÖ voters could be higher than those among others due to an increased share of unreported cases, i.e. a higher dark figure.

After finding some evidence in favor of the self-selection testing hypothesis in a positive correlation after the policy turn between the case fatality rate and the FPÖ vote share in 2019, I turned to individual-level data in order to further explore this hypothesis. I used the Austrian Covid-19 prevalence study conducted in November 2020 (Paškvan et al. 2021), which is a data source that contains the true infection status of individuals as determined by PCR tests, as well as one survey item regarding policy views. I found that "corona skeptics", here defined as people who think that the containment measures are exaggerated, are more likely to test positive for Covid-19, more likely to be an undetected prior and current case, but not significantly more likely to be an officially known case. All of these findings support the self-selection bias hypothesis.

Finally, I tested whether the proposed mechanism, i.e. the testing bias, can indeed explain the aggregate-level outcomes in a simple theoretical framework. In order to do so, I extended the classical SIRD to incorporate testing (with corresponding quarantine), heterogeneous behavior and heterogeneous mixing. This model is populated with two groups who may behave differently: the corona skeptics and the majority, where the former group has a lower propensity to get tested (which is empirically calibrated using the data from the Austrian Covid-19 prevalence study) and may have a higher basic reproduction number. I explored the properties of this model for three cases: a) homogenous mixing, b) proportionate mixing, and c) homophilic mixing.

My analysis showed that even a simple homogenous mixing setting allows for a situation in which an increase in the share of skeptics increases deaths, but has a non-linear inverted U-curve shaped impact on reported infections (i.e. increases them for lower shares of skeptics, but decreases them for higher ones). I further showed that homophily in mixing reduces the degree of non-linearity of the impact of an increase in the number of "skeptics" on cases and deaths. Homophily in mixing seems to be plausible, as a) Austrian government policy had a special focus on reducing the transmission in situations where homophily can be expected to be low (e.g. compulsory mask-wearing in shops, public transport and schools etc.), whereas b) large-scale protests organized by "corona skeptics" may have significantly increased the number of social contacts in particular among "skeptics" (see Lange and Monscheuer 2021 for some evidence on the German case).

Perfect homophily in mixing hence enables an increase in deaths without a change in reported cases as a special case. However, the simulations also show that the parameter space that allows for a relationship that may be statistically indistinguishable from this special case is much larger, in particular if the empirical distribution of cases is noisy. This result suggests that this mechanism may be relevant beyond the special case. Hence, a testing bias as it can be found in individual-level data can indeed explain the apparent aggregate-level paradox that the corona populist turn of the FPÖ seemed to have influenced the regional distributions of deaths (as the FPÖ vote share is correlated with deaths per capita after the policy switch), but not reported cases (as there is no statistically significant relationship with the FPÖ vote share).

The research presented in this paper can be extended in numerous ways. In particular, it seems promising to study corona populism and skepticism in a more complex model. One way to go would be to increase the complexity within the SIR framework. For instance, one could investigate the effects of dynamic policies depending on the number of infected, similar to what is proposed by Neuwirth et al. (2020), or even study optimal policies (e.g. Acemoglu et al. 2021; Alvarez et al. 2020; Bethune and Korinek 2020; Piguillem and Shi 2022) in the face of a non-compliant fraction of the population.

Another option would be to turn to a new class of models. Agent-based models such as the COVID-Town model (Mellacher 2020) are capable of modeling the spread of the virus via social networks and explicitly modeled heterogeneous agents, who can follow sophisticated behavioral rules. This level of analysis can be expected to be highly useful to better understand the impact and evolution of corona skepticism. For instance, it may make a big difference whether a corona skeptic faces many customers or is introverted and unemployed. However, this method can also help to better understand the emergence and dynamics of corona skepticism, e.g. by modeling heterogeneous risk perceptions, willingness to take risks, or even opinion dynamics of corona skepticism or corona populism.

More generally, it would be interesting to explore the relationship between mass opinion and party policy in a general theory of opinion dynamics (e.g. Hegselmann and Krause 2002; Mäs and Flache 2013; Mellacher 2021; Schweighofer et al. 2020) that incorporates both the "Downsian" effect and the "party identification" effect, hence complementing existing research in the Downsian tradition (e.g. Adams et al. 2005; Osborne 1993). I hope to be able to study some of these questions in the future.

**Declaration of interest:** None.

**Acknowledgements**: A previous version of this paper was featured in the preprint series "Covid Economics", Vol. 63. I want to especially thank Friederike Mengel for her clear editorial guidance and most helpful comments. I further thank five anonymous reviewers, Richard Sturn, Christian Neuwirth, Hans Manner, Golnaz Baradaran Motie, Tobias Eibinger, and the participants of i) the PhD seminar at the Graz doctoral school in economics, ii) the political science day 2021 of the Austrian Political Science Association, as well as iii) the research day 2022 of the School of Business, Economics and Social Sciences of the University of Graz for helpful suggestions, discussions and/or comments on previous versions of this paper. All remaining errors are mine. This research was funded in whole, or in part, by the Austrian Science Fund (FWF) [P 35228]. For the purpose of open access, the author has applied a CC BY public copyright licence to any Author Accepted Manuscript version arising from this submission.